\documentclass[useAMS, usenatbib]{mn2e}
\usepackage{graphicx, rotate}
\usepackage{natbib, lscape}
\usepackage{latexsym, amssymb, verbatim}
\usepackage[toc,page,header]{appendix}

\title[Characterisation of the molecular gas in ETGs]{The ATLAS$^{\rm
    3D}$ project - XVI. Physical parameters and spectral line energy
  distributions of the molecular gas in gas-rich early-type galaxies.}

\author[Bayet et al.]
  {Estelle Bayet,$^{1}$\thanks{E-mail: \texttt{bayet@physics.ox.ac.uk}}
  Martin Bureau,$^{1}$ 
  Timothy A.\ Davis,$^{2}$ 
  Lisa M.\ Young,$^{3}$ 
  \newauthor Alison F.\ Crocker,$^{4}$ 
  Katherine Alatalo,$^{5}$ 
  Leo Blitz,$^{5}$ 
  Maxime Bois,$^{6}$ 
  \newauthor Fr\'ed\'eric Bournaud,$^{7}$ 
  Michele Cappellari,$^{1}$ 
  Roger L.\ Davies,$^{1}$ 
  P.\ T.\ de Zeeuw,$^{2,8}$ 
  \newauthor Pierre-Alain Duc,$^{7}$ 
  Eric Emsellem,$^{2,9}$ 
  Sadegh Khochfar,$^{10}$
  Davor Krajnovi\'c,$^{2}$ 
  \newauthor Harald Kuntschner,$^{11}$ 
  Richard M.\ McDermid,$^{12}$ 
  Raffaella Morganti,$^{13,14}$ 
  \newauthor Thorsten Naab,$^{15}$ 
  Tom Oosterloo,$^{13,14}$ 
  Marc Sarzi,$^{16}$ 
  Nicholas Scott,$^{17}$ 
  \newauthor Paolo Serra,$^{13}$
  Anne-Marie Weijmans $^{18}$
  \vspace*{0.5cm}\\
  $^{1}$Sub-dept.\ of Astrophysics, Dept.\ of Physics, University of
  Oxford, Denys Wilkinson Building, Keble
  Road, Oxford, OX1 3RH, UK\\
  $^{2}$European Southern Observatory, Karl-Schwarzschild-Str. 2, 85748 Garching, Germany\\
$^{3}$Physics Department, New Mexico Institute of Mining and
Technology, Socorro, NM 87801, USA\\ 
$^{4}$University of Massachusetts, Amherst, USA\\
$^{5}$Department of Astronomy, Campbell Hall, University of
California, Berkeley, CA 94720, USA\\
$^{6}$Observatoire de Paris, LERMA and CNRS, 61 Av. de
l`Observatoire, F-75014 Paris, France\\
$^{7}$Laboratoire AIM Paris-Saclay, CEA/IRFU/SAp -- CNRS --
Universit\'e Paris Diderot, 91191 Gif-sur-Yvette Cedex, France\\
$^{8}$Sterrewacht Leiden, Leiden University, Postbus 9513, 2300 RA
Leiden, the Netherlands\\ $^{9}$Universit\'e Lyon 1, Observatoire
de Lyon, Centre de Recherche Astrophysique de Lyon and Ecole
Normale Sup\'erieure de Lyon, 9 avenue Charles Andr\'e, F-69230
Saint-Genis Laval, France\\ $^{10}$Max-Planck Institut f\"ur
extraterrestrische Physik, PO Box 1312, D-85478 Garching,
Germany\\
$^{11}$Space Telescope European Coordinating Facility, European
Southern Observatory, Karl-Schwarzschild-Str. 2, 85748 Garching,
Germany\\ $^{12}$Gemini Observatory, Northern Operations Centre,
670 N. A`ohoku Place, Hilo, HI 96720, USA\\ $^{13}$Netherlands
Institute for Radio Astronomy (ASTRON), Postbus 2, 7990 AA
Dwingeloo, The Netherlands\\ $^{14}$Kapteyn Astronomical
Institute, University of Groningen, Postbus 800, 9700 AV
Groningen, The Netherlands\\ $^{15}$Max-Planck-Institut f\"ur
Astrophysik, Karl-Schwarzschild-Str. 1, 85741 Garching,
Germany\\
$^{16}$Centre for Astrophysics Research, University of
Hertfordshire, Hatfield, Herts AL1 9AB, UK\\ $^{17}$Centre for
Astrophysics \& Supercomputing, Swinburne University of
Technology, PO Box 218, Hawthorn, VIC 3122, Australia
\\$^{18}$Dunlap fellow at the Dunlap Institute, University of Toronto, 50 St. George
Street, Toronto, ON M5S 3H4, Canada}

\date{Submitted to MNRAS on  ; Received ; in original form }

\pagerange{\pageref{firstpage}--\pageref{lastpage}} \pubyear{2012}

\begin{document}

\label{firstpage}
\maketitle

\clearpage
\begin{abstract}
  We present a detailed study of the physical properties of the
  molecular gas in a sample of $18$ molecular gas-rich early-type
  galaxies (ETGs) from the ATLAS$^{\rm 3D}$ sample. Our goal is to
  better understand the star formation processes occurring in those
  galaxies, starting here with the dense star-forming gas. We use
  existing integrated $^{12}$CO(1-0, 2-1), $^{13}$CO(1-0, 2-1),
  HCN(1-0) and HCO$^{+}$(1-0) observations and new $^{12}$CO(3-2)
  single-dish data. From these, we derive for the first time the
  average kinetic temperature, H$_{2}$ volume density and column
  density of the emitting gas in a significant sample of ETGs, this
  using a non-local thermodynamical equilibrium (non-LTE) theoretical
  model. Since the CO lines trace different physical conditions
  than of those the HCN and HCO$^{+}$ lines, the two sets of lines are
  treated separately. For most of the molecular gas-rich ETGs studied
  here, the CO transitions can be reproduced with kinetic temperatures
  of $10$-$20$~K, H$_{2}$ volume densities of $10^{3-4}$~cm$^{-3}$ and
  CO column densities of $10^{18-20}$~cm$^{-2}$. The physical
  conditions corresponding to the HCN and HCO$^{+}$ gas component have
  large uncertainties and must be considered as indicative only. We
  also compare for the first time the predicted CO spectral line
  energy distributions (SLEDs) and gas properties of our molecular
  gas-rich ETGs with those of a sample of nearby well-studied disc
  galaxies. The gas excitation conditions in $13$ of our $18$ ETGs
  appear analogous to those in the centre of the Milky Way, hence the
  star formation activity driving these conditions is likely of a
  similar strength and nature. Such results have never been obtained
  before for ETGs and open a new window to explore further star-formation
  processes in the Universe. The conclusions drawn should nevertheless
  be considered carefully, as they are based on a limited number of
  observations and on a simple model. In the near future, with higher CO
  transition observations, it should be possible to better identify
  the various gas components present in ETGs, as well as more
  precisely determine their associated physical conditions. To achieve
  these goals, we show here from our theoretical study, that mid-J CO lines (such as
  the $^{12}$CO(6-5) line) are particularly useful.
\end{abstract}


\section{Introduction}\label{sec:intro}

Early-type galaxies (ETGs) are central to our understanding of the
mass assembly history of galaxies. Historically, the SAURON
project\footnote{http://www.strw.leidenuniv.nl/sauron/} \citep{DeZe02}
aimed to understand the formation and evolution of elliptical
galaxies, lenticular galaxies and spiral bulges from three-dimensional
(3D) observations, this using the panoramic optical integral-field
spectrograph {\tt SAURON} \citep{Baco01}. However, a more
statistically robust sample was required to test its conclusions, and
more recently the ATLAS$^{\rm 3D}$
project\footnote{http://www-astro.physics.ox.ac.uk/atlas3d/} has
targeted a complete sample of $260$ early-type galaxies within the
local ($42$~Mpc) volume, also combining multi-wavelength surveys,
numerical simulations and semi-analytic modelling of galaxy formation
(\citealt{Cape11a}, hereafter Paper~I). ATLAS$^{\rm 3D}$ thus
  aims to constrain the dynamics of a complete volume-limited sample
  of ETGs, to fully characterise this class of galaxies and explore
  their formation and evolution.

The ATLAS$^{\rm 3D}$ dataset now provides an accurate inventory of the
baryon budget of local ETGs, with a detailed two-dimensional (2D)
description of the stellar and gaseous (both cold and warm)
kinematics, and resolved stellar population information for all
galaxies. It allows to probe the fossil record and ask a
  number of specific questions. How do slow rotators form? What are
  the physical processes determining their kinematic and photometric
  properties? What is the role of major and minor mergers in their
  formation history? How is star formation in ETGs quenched? Is the
  mechanism different for fast- and slow-rotating ETGs? How does it
  depend on environment? Etc. Earlier ATLAS$^{\rm 3D}$ papers have
already partially addressed many of these questions (e.g.\
\citealt{Bois10, Bois11}, hereafter Paper~VI; \citealt{Cape11b},
hereafter Paper~VII; \citealt{Davi11b}, hereafter Paper~X;
\citealt{Emse11}, hereafter Paper~III; \citealt{Khoc11}, hereafter
Paper~VIII; \citealt{Kraj11}, hereafter Paper~II; \citealt{Youn11},
hereafter Paper~IV).

ATLAS$^{\rm 3D}$ has a built-in effort to study the cold gas in ETGs,
both atomic (e.g.\ H{\small I}; \citealt{Morg06, Oost10, Serr12},
hereafter Paper~XIII) and
molecular (e.g.\ $^{12}$CO; \citealt{Comb07}; Paper~IV). These and
other efforts in the literature \citep[e.g.][]{Knap85, Kenn86, Knap96,
  Welc03, Sage06, Sage07} have been fuelled by the realisation over
the past few decades that many ETGs, previously thought to be `red and
dead', actually harbour a substantial amount of cold (hence molecular)
gas, and that many have a non-negligible level of star formation
\citep[e.g.][]{Welc03, Comb07, Croc11}.

More chemically complex species such as HCO$^{+}$ and HCN as well as
the less abundant isotopomer $^{13}$CO have also recently been
observed in ATLAS$^{\rm 3D}$ ETGs (\citealt{Krip10};
\citealt{Croc12}, hereafter Paper~XI), mostly using single-dish
telescopes. Paper~XI presents IRAM 30m telescope observations of the
$18$ $^{12}$CO-brightest ATLAS$^{\rm 3D}$ ETGs in the $^{13}$CO(1-0),
$^{13}$CO(2-1), HCN(1-0) and HCO$^+$(1-0) lines. Several key molecular
line ratios are investigated as empirical indicators of the molecular
gas physical conditions. Comparison with spiral galaxies shows that
the line ratios of ETGs and spirals generally overlap, but many ETG
outliers also exist. A few ETGs have more starburst-like line ratios,
while a few others have very low $^{12}$CO$/^{13}$CO
ratios. Correlations of line ratios involving $^{12}$CO(1-0) are found
with many galaxy properties: molecular-to-atomic gas mass fraction,
dust temperature, dust morphology and stellar population
age. Overall, it seems as if the driver of these correlations
  is a reduced optical depth in ETGs that have recently acquired or
  are in the process of acquiring their cold gas.

While these results are suggestive, Paper~XI does not include
modelling of the line ratios to derive the physical conditions of the
molecular gas. Detailed modelling has been performed for only one ETG
to date, Centaurus~A \citep{Ebne83, Ecka90a, Lisz01}, with several
molecular absorption lines observed towards its nucleus
\citep[e.g.][]{Ecka90a, Wild97, Mull09}. No systematic and homogenous
study, based on a comparison of observations of a well-defined sample
of ETGs with theoretical models, has ever been performed. To better
understand the evolution of the cold gas and associated star formation
in ETGs, it is however essential to constrain gas physical parameters
such as the kinetic temperature, H$_{2}$ volume density and the column
density of species such as CO, HCN and HCO$^{+}$, even if they are
averaged over entire galaxies.

Here, we therefore aim to estimate the globally-averaged physical
properties of the molecular gas in a significant sample of ETGs
($18$), using a single-dish dataset (largely the observations
presented in Paper~XI) and focussing on the two gas components traced
by the CO and HCN/HCO$^{+}$ lines, respectively. As this first study
is by necessity exploratory, we use a simple theoretical framework,
non-local thermodynamical equilibrium (non-LTE) model, specifically
the Large Velocity Gradient (LVG) model developed by
\citet{VanderTak07}. We have also recently obtained $^{12}$CO(3-2)
observations with the HARP instrument on the James Clerk Maxwell
Telescope (JCMT) for a few objects. We report both on these new
observations and the theoretical model in this paper.

The structure of the paper is as follows. We first present new
  CO(3-2) observations of $8$ sample galaxies with existing multi-line
  observations in Section~\ref{sec:obs}, and briefly review the main
characteristics of the other observations on which our theoretical
analysis is based. The main assumptions and parameters of the LVG
model are described in Section~\ref{sec:mod}. The comparison of the
model predictions and observations, and the identification of the
models reproducing best the data are detailed in
Section~\ref{sec:resu}. A discussion of our results and literature
data are then presented in Section~\ref{sec:disc}. The results are
discussed further in Section~\ref{sec:con} where we also conclude
briefly. We remind the reader that all the basic properties of
  our sample galaxies can be found in previous ATLAS$^{\rm 3D}$ papers
  (e.g.\ Paper~I).

\section{Observations}\label{sec:obs}

\subsection{Literature data}\label{subsec:lit}

Most of the $^{12}$CO, $^{13}$CO, HCN and HCO$^{+}$ molecular gas
detections in ETGs that we are using here for modelling are taken from
previously published SAURON and ATLAS$^{\rm 3D}$ follow-ups.
Observational parameters beyond those detailed below can be found in
these studies (\citealt{Welc03, Comb07, Krip10}; ; Paper~IV; Paper~XI). To
be more precise, observations of the HCN(1-0) ($\nu=89.087$~GHz),
HCO$^{+}$(1-0) ($\nu=89.188$~GHz), $^{13}$CO(1-0) ($\nu=110.201$~GHz)
and $^{13}$CO(2-1) ($\nu=220.398$~GHz) lines were presented in \citet{Krip10} and
Paper~XI, following-up on the $18$ strongest $^{12}$CO(1-0)
($\nu=115.271$~GHz) and $^{12}$CO(2-1) ($\nu=230.538$~GHz) detections
of \citet{Comb07} and Paper~IV.

\subsection{New $^{12}$CO(3-2) observations}\label{subsec:CO3-2}

\begin{figure*}
    \centering
    \includegraphics[scale=0.6]{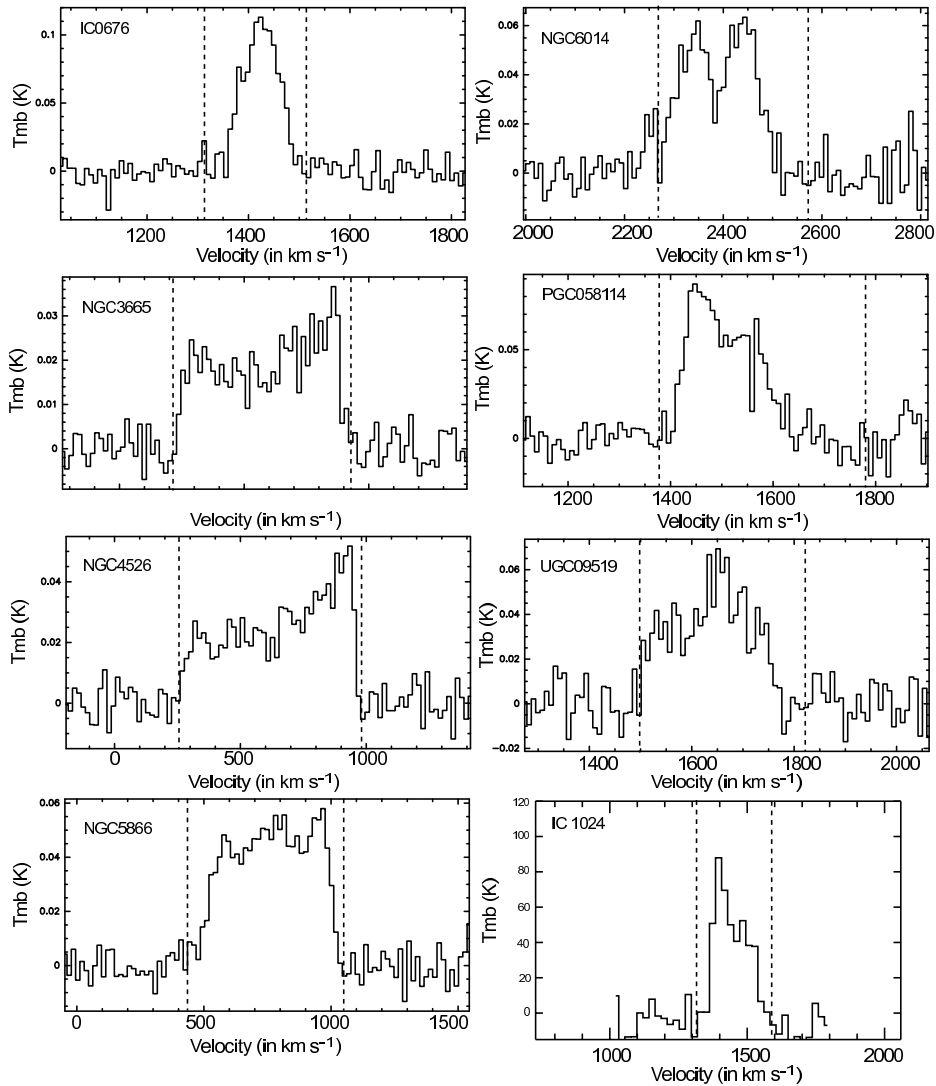}
    \caption{New JCMT $^{12}$CO(3-2) line observations of $8$ galaxies
      from our sample. As the line profiles are not always Gaussian,
      velocity-integrated line intensities were extracted by simply
      summing the intensities over the range of velocities (indicated
      by the dashed lines) with detected $^{12}$CO(1-0) emission (see
      Paper~IV). Although IC1024 is clearly detected in
        $^{12}$CO(3-2), both the shape and scaling of its line profile
        are highly uncertain due to a pointing error. We thus present
        the IC1024 data here for completeness only.}
\label{fig:0}
\end{figure*}

New $^{12}$CO(3-2) ($\nu=345.795$~GHz) observations of $8$ of
  those $18$ galaxies were acquired at the JCMT (project m11au11)
between 31 March and 4 September 2011. These galaxies are
  pathfinders for a larger $^{12}$CO(3-2) survey, and were selected
  because of their bright $^{12}$CO(1-0) fluxes. For $5$ galaxies
(IC0676, IC1024, NGC6014, PGC058114 and UGC09519), the HARP receiver
and ACSIS backend were used in a configuration yielding a $1$~GHz
total bandwidth and $2048$ channels each $488$~kHz-wide. Three
galaxies (NGC3665, NGC4526 and NGC5866) have broader lines (see
Fig.~\ref{fig:0}) and for those the backend configuration used two
overlapping sub-bands, each $1$~GHz wide and covering a total
bandwidth of $1.9$~GHz with $977$~kHz-wide channels. Observations were
carried out in the grid-chop or staring mode, with beam switching and
a $90\arcsec$ throw. Opacities at $225$~GHz ($\tau_{225}$) ranged from
$0.05$ to $0.1$ and system temperatures (at $345$~GHz) from $230$ to
$370$~K. Pointing tests before each galaxy observation and less
frequent focus calibrations were made on the standard sources RXBoo,
IRC+10216, IRC+20326, CIT6, RTVir, and XHer, and the average pointing
offset was 3$''$-4$''$. Repeated observations of the standard sources
G34.3 and 16293-2422 every night suggest that the absolute calibration
is constant at the $2$ to $5\%$ level. Stable baselines required only
a $0^{\rm th}$ order baseline subtraction. Conversion from antenna
temperature to main beam brightness temperature was done via $T_{\rm
  mb}=T_{\rm A}^*/0.6$ (see \citealt{Warr10} or the references in
http://www.jach.hawaii.edu/JCMT/spectral\_line/ and
http://www.jach.hawaii.edu/JCMT/spectral\_line/Standards /current\_cals.html). The
half-power beam width (HPBW) of the JCMT at $\nu\approx345$~GHz is
$14\arcsec$.

The data pre-reduction for the new $^{12}$CO(3-2) observations was
done using Starlink software (KAPPA, SMURF and STLCONVERT packages\footnote{See
  http://www.jach.hawaii.edu/JCMT/spectral\_line/ data\_reduction/acsisdr/basics.html}),
and the data were subsequently translated to CLASS\footnote{See http://www.iram.fr/IRAMFR/GILDAS/doc /html/class-html/class.html} format for final
reduction. The reduced spectra for all $8$ galaxies are
presented in Figure~\ref{fig:0} and the integrated intensities are
listed in Table~\ref{tab:0}. No other CO(3-2) observation is
  available so far. The $^{12}$CO(3-2) line profile in NGC4526 is
asymmetric, probably caused by a combination of a small pointing error
and the emission region being slightly larger than the beam at this
frequency (causing some ``missing'' flux at low velocities). With only
single-dish information, it is difficult to precisely estimate how
much flux is missed. However, we do have recently acquired
  interferometric $^{12}$CO(2-1) CARMA data of NGC4526 (Davis et al.\, in prep.), and estimate $\approx8\%$ of
  missing flux. We have checked that our conclusions
remain unchanged when using the corrected flux (see
Section~\ref{sec:resu}). The $^{12}$CO(3-2) line emission of
  IC1024 suffers from a greater pointing uncertainty, that does not
  allow us to use this detection in our modelling work
  (Section~\ref{sec:mod}). Observations of IC1024 alone used the
  jiggle-chop mode with a $4\times4$ map-centered grid, yielding a
  square grid of spectra spaced by $7\farcs5$. The optical center of
  the galaxy therefore does not fall on a grid position, and we
  averaged the four central spectra to produce the spectrum shown in
  Figure~\ref{fig:0}. The resulting averaged line profile is however
  significantly different from those of the $^{12}$CO(1-0) and
  $^{12}$CO(2-1) lines (see Paper~XI), with a peak flux velocity more
  than $100$~km~s$^{-1}$ away from those of the lower-J CO lines and a
  $^{12}$CO(3-2) line width $20\%$ smaller. Any $^{12}$CO(3-2) flux
  correction would thus be highly uncertain, and we do not include
  this detection in our modelling work. Overall, we therefore have
reliable $^{12}$CO(3-2) observations for $7$ of our $18$ sample
galaxies.

\begin{table}
  \caption{Velocity-integrated line intensities of
    the new $^{12}$CO(3-2) JCMT data (see Section~\ref{sec:obs}).}
\label{tab:0}
\begin{center}
\begin{tabular}{lr}
\hline
Galaxy    & $\int{T_{\rm mb}\,\Delta v}$\\
          & (K~km~s$^{-1}$)\\
\hline
IC0676    & 10.14$\pm 0.34$\\
IC1024    & 9.57$\pm 0.74$\\
NGC3665   & 12.32$\pm 0.41$\\
NGC4526   & 18.58$\pm 0.58$\\
NGC5866   & 23.84$\pm 0.59$\\
NGC6014   & 8.60$\pm 0.46$\\
PGC058114 & 9.01$\pm 0.47$\\
UGC09519  & 11.93$\pm 0.52$\\
\hline
\end{tabular}
\end{center}
\end{table}

\subsection{Data reduction and analysis}\label{subsec:reduction}

The data reduction for all galaxies, including the
  $^{12}$CO(3-2) line, has been performed (or redone when already
published) in a systematic and homogeneous way using the GILDAS
package, as described in Paper~XI. As the line profiles are not always
Gaussian, velocity-integrated line intensities were extracted by
simply summing the intensities over the range of velocities with
detected $^{12}$CO(1-0) emission (see Paper~IV). Although the
  $^{12}$CO(3-2) emission of NGC6014 and PGC058114 is slightly
  off-centred within that range, we have kept it for consistency with
  previously published results. The statistical uncertainties
resulting from this method are discussed in great detail in Paper~XI
and lead to errors on the velocity-integrated line intensities of
$\leqslant10\%$. If the peak flux is lower than three times the rms
noise in line-free channels (integrated
over the expected velocity range), as is the case for several galaxies in
the HCN(1-0) and HCO$^{+}$(1-0) lines, the line is considered
undetected and an upper limit of three times the rms noise is
adopted.

The LVG models (see Section~\ref{subsec:LVG}) also require the
observed line widths of each transition and molecule. Depending on the
source and spectrum, we use either a Gaussian or a double-peaked line
profile fit to extract the line width, taking the line width as the
full-width at half maximum (FWHM) for Gaussian profiles and the
velocity width given by the shell method for double-peaked profiles
\footnote{More details are avaible in the GILDAS package for both
  methods.}. These provide satisfactory estimates given the crude
model assumptions. The line widths are tabulated in Table~\ref{tab:1},
where the errors reported are formal fitting uncertainties, with an
average of $20\%$. For undetected lines (flux upper limits), we take a
line width equal to that of the 'closest' molecular species. For
example, when undetected, we have assumed that HCO$^{+}$(1-0) has the
same line width as the HCN(1-0), since we consider that both species
trace the same gas component. Similarly, when the $^{12}$CO(2-1) or
$^{13}$CO(2-1) transitions only have upper limits, their line widths
are assumed to be the same as those of the $^{12}$CO(1-0) and
$^{13}$CO(1-0) transitions, respectively.  When both HCN and HCO$^{+}$
are undetected, we have not attempted to calculate the corresponding
line ratio nor to identify any best model, as the ratio of two upper
limits is unconstrained. For those cases, we do not provide any line
width either (see Table~\ref{tab:1}).

\section{Non-LTE Models}\label{sec:mod}

\subsection{Generalities}\label{subsec:general}

As shown theoretically \citep[e.g.][]{Meij07, Spaa05, Baye08a,
  Baye09a} and confirmed observationally \citep[e.g.][]{Isra03,
  Gao04b, Kram05, Garc06, Baye08b, Baye09c, Alad10}, there are
several distinct cold gas components in external galaxies, each traced
by specific molecular tracers. For our modeling, we thus consider a
simple approach with two gas components: one traced by the CO lines (a
rather extended and cool phase) and another traced by the HCN and
HCO$^{+}$ lines (expected to be warmer and denser). This approach is
most appropriate given the significant differences between the
critical line densities of these species, i.e.\ more than two orders
of magnitude between HCO$^{+}$(1-0) and $^{12}$CO(1-0), and three
orders of magnitude between HCN(1-0) and $^{12}$CO(1-0). The
  gas traced by the low-J CO lines and the gas traced by HCO$^{+}$
and HCN are thus not compatible; mixing these two components would not
allow us to increase our knowledge of the interstellar medium (ISM)
properties of ETGs as much as the two-component approach
adopted. We do note that HCN(1-0) and HCO$^{+}$(1-0) do not
  have exactly the same critical line density ($\approx1$ order of
  magnitude difference), and thus they do not strictly speaking trace
  the same gas component. However, in the rest of the current paper we
  will consider that they do, in the sense that they both
  qualitatively trace a gas component denser than that observed with
  the low-J CO lines.

In the following, we refrain to estimate gas masses and gas
  fractions, as there are large uncertainties in the values of the
  appropriate conversion factors for ETGs (CO-to-H$_{2}$ and
  HCN-to-H$_{2}$). In fact, there is currently no estimate of the
  HCO$^{+}$-to-H$_{2}$ conversion factor at all in the literature,
  even for late-type galaxies, so it is clearly not possible to
  estimate the gas mass and gas fraction associated with this
  species. Nevertheless, qualitative empirical estimates of gas
  fractions based on the line ratios of $^{12}$CO, $^{13}$CO, HCN and
  HCO$^{+}$ were discussed in Paper~XI.

\subsection{Model description}\label{subsec:LVG}

To determine the average physical properties of the CO gas component
on the one hand, and those of the HCN and HCO$^{+}$ gas component on
the other hand, we use RADEX, the non-LTE model developed by
\citet{VanderTak07}. This model is a zero-dimensional, one molecular
species code. It solves the radiative transfer equations by assuming a
homogeneous medium and retaining the assumption of local
excitation. To partially decouple the level population and radiative
transfer equations, the \citet{VanderTak07} model uses the escape
probability method (one variation of which is the LVG method) assuming
a uniform (i.e.\ non-expanding) sphere. The escape probability
equation is then given by \citet{Oste74}. The model also assumes that
the optical depth is independent of velocity, appropriate for
modelling velocity-integrated line intensities.

Although the assumption of uniform physical conditions for each gas
component over the scale of a galaxy is crude (e.g.\ the ISM of nearby
galaxies is known to be clumpy), these models do provide first order
constraints on the average physical conditions, and they do represent
a useful first step in predicting the intensities of the higher order
transitions of CO, HCN and HCO$^{+}$, on which there is currently no
constraint in ETGs. In any case, we currently lack information on the
very small-scale ($< 70$~pc) gas distribution of ETGs, preventing
us from studying their gas properties in greater detail.

Molecular collisional rates are required to derive the molecular gas
physical conditions using this model, in addition to the spectroscopic
and dipole moment data\footnote{From The Cologne Database for
  Molecular Spectroscopy; see http://www.astro.uni-koeln.de/cdms/~.}
needed when using more basic (e.g.\ LTE) models. We used collisional
rates for CO, HCN and HCO$^{+}$ from the Leiden Atomic and Molecular
Database
(LAMDA)\footnote{http://www.strw.leidenuniv.nl/$\sim$moldata/} as
recommended in \citet{VanderTak07}.

There are five main parameters in the LVG model we use: the column
density $N(X)$ of the species studied $X$ (here either $^{12}$CO,
$^{13}$CO, HCN or HCO$^{+}$) in cm$^{-2}$, the line width $\Delta v$
of each transition (see Table~\ref{tab:1}) in km~s$^{-1}$, the
molecular hydrogen volume density $n$(H$_{2}$) in cm$^{-3}$, the
kinetic temperature $T_{\rm K}$ in K and the abundance ratios (e.g.\
isotopic ratios such as $^{12}$C$/^{13}$C and the HCN/HCO$^{+}$
abundance ratio). In our work, we leave the three parameters
  $T_{\rm K}$, $n$(H$_{2}$) and $N(X)$ free (and solve for them by
  fitting the data), whereas we fix the other parameters ($\Delta v$
  and isotopic abundance ratios). We note that we do not implement
more specific characteristics of ETGs in our LVG models, such as high
metallicity or enhanced $\alpha$-element abundances. These
particularities and their detailed influence on the molecular
chemistry of ETGs have however been considered in \citet{Baye12a}.

\subsection{Choice of parameters}\label{subsec:param}

As discussed above (Section~\ref{subsec:reduction}), the fixed
parameter $\Delta v$ is taken as the FWHM of the fit for Gaussian
profiles or the line width of the fit for double-peaked profiles, and
is measured for each molecular line in each source (see
Table~\ref{tab:1}). It varies from $58$ to $425$~km~s$^{-1}$, and its
value directly impacts on the predicted velocity-integrated line
intensity calculated by the LVG model. Indeed, the predicted
velocity-integrated intensities are obtained by multiplying the
calculated radiation temperature of the spectral line by $1.06\,\Delta
v$, where the factor $1.06$ is due to the assumption of a Gaussian
line profile. Although not all line profiles are Gaussian, the models
are constrained only by ratios of velocity-integrated line
intensities, so this factor and the assumed shape of the line profiles
have no influence on the model fits; only variations in the line
profiles across the
different lines involved in the ratios matter. This is no longer true for the calculation of the
predicted spectral line energy distributions (SLEDs), however, and the
SLEDs of galaxies with non-Gaussian line profiles are therefore more
uncertain (see Section~\ref{subsec:SED}).

\begin{table*}
  \caption{Line widths for our sample of 18 gas-rich ETGs, extracted
    from the observations published in Paper~XI and from the new $^{12}$CO(3-2)
    observations presented here.}
  \label{tab:1}
  \hspace*{-1.5cm}
  \resizebox{21cm}{!}{
    \begin{tabular}{lcrrrrrrr}
      \hline
      Galaxy & D &  $\Delta v\,(^{12}$CO(1-0)) & $\Delta
      v\,(^{12}$CO(2-1)) &  $\Delta v\,(^{12}$CO(3-2)) & $\Delta v\,(^{13}$CO(1-0)) & $\Delta
      v\,(^{13}$CO(2-1)) & $\Delta v\,$(HCN(1-0)) & $\Delta v\,$(HCO$^{+}$(1-0)) \\
      & (Mpc) & (km~s$^{-1}$) & (km~s$^{-1}$)& (km~s$^{-1}$)& (km~s$^{-1}$)& (km~s$^{-1}$)& (km~s$^{-1}$)&(km~s$^{-1}$)\\
      \hline
      IC0676     & 24.6 & $97.3\pm17.4$  & $87.1\pm10.7$ & $84.1\pm\phantom{0}3.1$ & $150.2\pm39.1$ & $77.7\pm16.1$  & $101.2\pm28.1$  & $116.8\pm26.7$  \\
      IC1024     & 24.2 & $160.9\pm19.3$ & $146.9\pm22.4$ & $123.1 \pm 20.0$ &$191.0\pm30.6$ & $152.6\pm19.8$ & -               & -               \\
      NGC1222    & 33.3 & $158.7\pm10.4$ & $169.8\pm15.7$ & - & $170.8\pm31.8$ & $133.7\pm17.7$ & -               & $128.7\pm30.0$  \\
      NGC1266    & 29.9 & $128.1\pm22.3$ & $152.2\pm28.9$ &  - &$95.1\pm19.1$  & $104.0\pm20.5$ & $157.3\pm48.3$  & $179.6\pm13.7$  \\
      NGC2764    & 39.6 & $249.5\pm\phantom{0}6.6$  & $226.0\pm\phantom{0}7.0$  & - &$250.9\pm22.8$ & $246.3\pm25.0$ & $220.3\pm62.4$  & $199.2\pm34.6$  \\
      NGC3032    & 21.4 & $128.8\pm14.5$ & $102.5\pm\phantom{0}3.3$  & - &$80.8\pm16.5$  & $103.4\pm11.2$ & $162.5\pm45.5$  & -               \\
      NGC3607    & 22.2 & $268.6\pm40.5$ & $323.1\pm\phantom{0}4.7$  &  - &$425.9\pm85.2$ & $386.7\pm68.6$ & $385.4\pm71.4$  & $387.3\pm$77.8  \\
      NGC3665    & 33.1 & $318.1\pm64.7$ & $319.4\pm51.7$ & $317.8\pm54.7$&$327.5\pm65.5$ & $312.6\pm76.8$ & $318.3\pm27.8$  & -               \\
      NGC4150    & 13.4 & $159.2\pm13.8$ & $189.1\pm17.6$ & - &$149.0\pm29.8$ & $158.2\pm30.4$ & -               & -               \\
      NGC4459    & 16.1 & $194.6\pm35.7$ & $192.7\pm28.9$ & - &$196.3\pm33.3$ & $190.1\pm39.3$ & $193.3\pm42.5$  & -               \\
      NGC4526    & 16.4 & $348.9\pm34.9$ & $341.1\pm68.0$ &$336.7\pm53.8$ &$345.6\pm55.3$ & $345.2\pm72.5$ & 346.7$\pm73.4$  & -               \\
      NGC4694    & 16.5 & $80.2\pm\phantom{0}5.2$   & $70.9\pm14.2$  &  - &$57.9\pm\phantom{0}9.3$   & $100.8\pm30.3$ & -               & -               \\
      NGC4710    & 16.5 & $229.2\pm45.3$ & $209.5\pm42.0$ & - &$292.7\pm52.8$ & $261.8\pm52.1$ & $243.0\pm59.5$  & $272.5\pm43.4$  \\
      NGC5866    & 14.9 & $294.8\pm50.5$ & $242.5\pm 48.6$& $240.7\pm43.3$&$374.3\pm71.1$ & $377.9\pm74.1$ & $338.6\pm71.9$  & $360.5\pm33.6$  \\
      NGC6014    & 35.8 & $179.7\pm28.8$ & $122.1\pm24.6$ & $107.2\pm14.0$&$162.1\pm25.9$ & $140.8\pm29.5$ & $93.3\pm17.9$   & -               \\
      NGC7465    & 29.3 & $129.8\pm24.8$ & $80.1\pm12.7$  & - &$140.8\pm19.7$ & $130.9\pm27.6$ & $150.5\pm42.6$  & $124.7\pm12.9$  \\
      PGC058114   & 23.8 & $155.8\pm29.6$ & $160.5\pm32.1$ &$149.4\pm22.4$& $174.9\pm27.9$ & $154.7\pm32.5$ & -               & $170.9\pm$54.4  \\
      UGC09519   & 27.6 &$ 179.5\pm32.1$ & $155.0\pm31.5$ & $128.0\pm15.7$&$153.1\pm26.0$ & $199.5\pm46.9$ & -               & $141.2\pm$34.2  \\
      \hline
    \end{tabular}}
  
  {\bf Notes:} The symbol `-' implies that no line width could be
  measured, either because there is no observation (e.g.\
  $^{12}$CO(3-2) line) or because signal-to-noise ratio of the line is
  too low. However, when required, the line width adopted
  is that from the `closest' species (see Section~\ref{sec:obs}).
\end{table*}

Currently, there is little knowledge of the true isotopic abundance
ratio of carbon in ETGs and of its variation from source to source. We
have thus assumed that it is constant across our sample. As explained
for instance by \citet{Mart10} and \citet{Robe11}, velocity-integrated
line intensity ratios can be misleading and are not a good proxy for
abundance ratios, as they are affected by geometry and optical depth
effects. The velocity-integrated line intensity ratios of
$^{12}$CO$/^{13}$CO provided in Paper~XI (see their Figure~3) can not
therefore be used to derive an accurate value of the intrinsic
$^{12}$C$/^{13}$C abundance ratio. Similarly for
HCN(1-0)/HCO$^{+}$(1-0). \citet{Wils94} reported values of the
$^{12}$C$/^{13}$C abundance ratio in the centre of the Milky Way, the
$4$~kpc molecular ring, the local ISM and the Solar system of
$\approx20$, $50$, $70$ and $90$, respectively. For local starburst
galaxies, it has been recently estimated to be $>46$
\citep{Mart10}. Based on optically thin lines, \citet{Robe11} reported
a HCN/HCO$^{+}$ abundance ratio between $0.5$ and $1.03$ for local and
extragalactic active star-forming regions. For the Galactic centre,
only optically thick lines of HCN and HCO$^{+}$ are present in the
literature (and are thus misleading). Lacking better information for
either case, we have therefore adopted a fixed value of $70$ for the
$^{12}$C$/^{13}$C abundance ratio and a value of $1$ for
HCN/HCO$^{+}$. Only knowledge of the small-scale distribution of the
$^{12}$CO, $^{13}$CO, HCN and HCO$^{+}$ emission for various line
transitions would allow us to properly estimate the true abundance
ratios.

We thus built four grids of LVG models to be combined, one for each of
$^{12}$CO, $^{13}$CO, HCN and HCO$^{+}$. Each grid covers kinetic
temperatures of $10$ to $250$~K in steps of $10$~K, gas volume 
densities of $10^{3}$ to $10^{7}$ cm$^{-3}$ in steps of $0.5$~dex,
column densities of $10^{10}$ to $10^{20}$ cm$^{-2}$ in steps of
$0.5$~dex and line widths of $60$ to $430$~km~s$^{-1}$ in steps of
$10$~km~s$^{-1}$. Each grid thus contains over $150,000$ models,
leading to a total of over $600,000$ models for the species
investigated in the present study.

The model outputs are numerous, and among them is the quantity we are
primarily interested in, i.e.\ the predicted velocity-integrated line
intensity (in K~km~s$^{-1}$). We ran each model to include the first
$15$ energy levels and therefore the first $15$ transitions of each
species, allowing us to predict the most interesting observational
lines for Herschel Observatory and the Atacama Large
Millimeter/sub-millimeter Array (ALMA) (see Section~\ref{subsec:SED}).

For each galaxy, we have thus compared the predicted
velocity-integrated line intensity ratios (computed directly from the
model outputs) with the observed ones. More specifically, we have
worked with the independent $^{12}$CO(1-0)/$^{13}$CO(1-0),
$^{12}$CO(2-1)/$^{13}$CO(2-1), $^{12}$CO(1-0)/$^{12}$CO(2-1) and
$^{12}$CO(1-0)/$^{12}$CO(3-2) (whenever the $^{12}$CO(3-2) line was
available; see Table~\ref{tab:0}) line ratios on the one hand, and the
HCN(1-0)/HCO$^{+}$(1-0) line ratio on the other hand. For each ratio,
we have corrected for beam dilution as described in Paper~XI. The
$^{12}$CO(3-2) line emission is only used in one ratio, not two,
because we wanted to keep independent constraints for reliable
$\chi^{2}$ calculations (see Section~\ref{subsec:likely}).

To identify the model whose input parameters ($T_{\rm K}$,
$n$(H$_{2}$) and $N(X)$) reproduce best the observed line ratios, we
have therefore constrained the models with a minimum of three observed
line ratios for the CO gas component, and only one observed line ratio
for the HCN and HCO$^{+}$ gas component. The physical properties
derived for the CO gas component are thus more reliable than those
derived for the HCN and HCO$^{+}$ one, as the latter are
under-constrained. The results for the HCN and HCO$^{+}$ gas component
should therefore be considered purely as indicative, and we present
these results (table and figures) in Appendix~\ref{app:fig} only. We
will focus exclusively on the CO gas component for the remainder of the
current paper.

Finally, we note that the observed line ratios involving isotopologues
($^{12}$CO(1-0)/$^{13}$CO(1-0) and $^{12}$CO(2-1)/$^{13}$CO(2-1)) have
been reproduced by combining a $^{12}$CO model having a column density
$N$($^{12}$CO) with a $^{13}$CO model having a column density
$N$($^{13}$CO)=($^{13}$C/$^{12}$C)$N$($^{12}$CO). Similarly, an
observed ratio line involving a $^{12}$CO(1-0) line with a FWHM of
$300$~km~s$^{-1}$ and a $^{12}$CO(2-1) line with a FWHM of
$290$~km~s$^{-1}$ has been reproduced by combining a $^{12}$CO model
having a velocity width of $300$~km~s$^{-1}$ for the (1-0) transition
with a $^{12}$CO model having a velocity width of $290$~km~s$^{-1}$
for the (2-1) transition.

\section{Results}\label{sec:resu}

We explore below two complementary methods to identify the `best'
model for each galaxy studied: the best-fit model based on the
$\chi^{2}$ method (see Section~\ref{subsec:chi}) and the most likely
model based on the likelihood (see Section~\ref{subsec:likely}). From
these two methods, we calculate for the first time the corresponding
predicted spectral line energy distributions of CO in local ETGs (see
Section~\ref{subsec:SED}).

\subsection{Best-fit model identification}\label{subsec:chi}

For each source, we identify a best model using the $\chi^2$
minimisation method for each of the two gas components investigated,
i.e.\ the gas traced by CO on the one hand and the gas traced
  by HCN and HCO$^{+}$ on the other. The physical properties ($T_{\rm
    K}$, $n$(H$_{2}$) and $N(X)$) corresponding to these best fits are
  listed respectively in Table~\ref{tab:2} for the CO gas component
  and in Table~\ref{tab:A1} for the HCN and HCO$^{+}$ component. This
is the first time that the physical properties of the molecular gas
are constrained in ETGs in a systematic and homogeneous way, and for a
relatively large number of sources ($18$). These properties are of
prime interest for further individual and statistical studies of the
gas in ETGs (physical conditions, origin, H{\small I}-to-H$_{2}$
ratio, etc). When calculating the $\chi^{2}$, the errors on the
observed line ratios have been properly calculated and are fully taken
into account. As the errors on line ratios are asymmetric (see
Paper~XI), we have considered the geometric average of the two values
in our $\chi^{2}$ calculations.

The kinetic temperatures obtained for the best-fit models are either
low ($T_{\rm K}=10$-$20$~K for IC1024, NGC1222, NGC1266, NGC2764,
NGC3032, NGC3665, NGC4150, NGC4459, NGC4526, NGC4694, NGC4710 and
NGC5866), intermediate ($30$-$70$~K for IC0676, NGC3607, PGC058114 and
UGC09519) or high ($120$~K for NGC6014 and $150$~K for
NGC7465). Although these kinematic temperatures may seem high
  in view of the commonly assumed low star-formation activity of ETGs,
  the fact is that the absolute star-formation rates (SFRs) of ETGs
  are not always low (especially in the molecular gas-rich objects
  considered here). It is the specific SFRs (SSFRs; i.e.\ the SFRs per
  unit stellar mass) that are generally lower than those of spirals
  (see, e.g., \citealt{Shap10} for the SAURON ETGs). ETGs also have
  additional heating, excitation and ionisation mechanisms that are
  unlikely to be important in other galaxy types. For example, old
  (and therefore not massive) but hot stars will heat the dust grains
  and, through the photoelectric effect, the gas, impacting both the
  molecular and ionised gas emission lines. This peculiar population
  also gives rise to the ultraviolet (UV)-upturn phenomenon
  \citep[e.g.][]{OCon99,Bure11,Jeon12}, increasing the UV flux (and
  thus the heating) surrounding the gas independently of star
  formation. Supernova rates leading to cosmic-ray fluxes may also be
  different than those of spirals. Finally, the high stellar
  metallicity and $\alpha$-element enhancement may also play a role in
  the gas chemistry \citep{Baye12a}. Any or all of these mechanisms
  may thus partly be responsible for the high gas temperatures
  measured in some galaxies, although it is impossible at this point
  to determine which has the greatest influence.

The best-fit H$_{2}$ volume densities do not vary by more than a
factor of $10$ between our sources
($n$(H$_{2}$)$=10^{3-4}$~cm$^{-3}$), except for NGC1222 and NGC5866
that have $n$(H$_{2}$)$=10^{6.5}$ and $10^{7}$~cm$^{-3}$,
respectively. Similarly, the CO column densities derived are pretty
constant amongst all the galaxies studied, with
$N$(CO)$=10^{18.5-20}$~cm$^{-2}$. These column densities may
  appear high, especially once converted to N(H$_{2}$). Indeed,
  assuming a standard Milky Way CO-to-H$_{2}$ abundance ratio of
  $10^{-4}$, one obtains N(H$_{2}$)$=10^{22.5-24}$~cm$^{-2}$,
  coresponding to an average surface mass density of
  $10^{2.5-4}$~$M_{\odot}$~pc$^{-2}$. For comparison, giant molecular
  clouds (GMCs) in the Milky Way have typical surface densities of
  $10^{2}$~$M_{\odot}$~pc$^{-2}$. However, as shown by \citet{Baye12a}
  who studied in details the chemistry of ETG-like environments, the
  standard Milky Way CO/H$_{2}$ fractional abundance is not
  appropriate for ETGs, and a value higher than
  $10^{-4}$ should be used (see Fig.~2 in \citealt{Baye12a}). For most
  of the sources studied here, a value of $10^{-2.5}$ is more
  appropriate, leading to surface densities closer to the Milky Way
  value. In addition, high CO column densities may not be totally
  unexpected, as suggested by our CO interferometric maps
  (\citealt{Alat13}, hereafter Paper~XVIII), where high surface
  densities are observed despite low SSFRs (as high as
  $\approx10^{4}$~$M_{\odot}~$pc$^{-2}$ in NGC1266). These CO column
  densities may also be considered as lower limits, since the authors
  conservatively assumed a standard CO-to-H$_{2}$ conversion
  factor. This apparent low star-formation efficiency (SFE; i.e.\ the
  star formation rate per unit dense gas mass) may partly be due to
  the different dynamics of ETGs compared to that of spirals
  (morphological quenching; see \citealt{Mart09}). Nevertheless, most
  ETGs show typical surface densities of
  $\approx10^{2}$~$M_{\odot}$~pc$^{-22}$, consistent with Milky Way
  values.

Table~\ref{tab:2} reveals that the galaxies detected in $^{12}$CO(3-2)
are not systematically those showing the highest kinetic temperatures
and gas densities as determined by the best-fit method (see, e.g.,
NGC3665 and NGC4526). As it takes warmer and denser gas to effectively
populate the (3-2) level, this may seem surprising at first, but in
fact our best-fit results only show that the $^{12}$CO(3-2) line in
these galaxies is not the main driver of our best-fit model
identification. In other words, there is basically nothing special about 
the objects detected in $^{12}$CO(3-2). In fact, we did not select for
observations objects predicted to have the highest absolute
$^{12}$CO(3-2) fluxes, but rather the highest {\em apparent}
$^{12}$CO(3-2) fluxes (e.g.\ because they are nearby or have high
$^{12}$CO(1-0) fluxes). So the other objects are not so much
non-detections as they are non-observations (we just do not know what
their $^{12}$CO(3-2) fluxes are, and they may well be high). A larger
sample of $^{12}$CO(3-2) observations in ETGs is thus needed for
further comparison.

Understanding the exact driver of the best-fit model identification
requires a detailed analysis of the LVG code outputs and of the
$\chi^{2}$ minimisation, that are outside the scope of this
paper. However, to evaluate the accuracy of our best-fit CO models and
calculate uncertainties on the derived best-fit model parameters
listed in Table~\ref{tab:2}, we have calculated the
$\Delta\chi^2\equiv\chi^2-\chi^2_{\rm min}$ contours for each source
in our 3D parameter space ($T_{\rm K}$, $n$(H$_{2}$),
$N$(CO)). The first five contours in Figure~\ref{fig:1} show the $1$ to
$5\sigma$\footnote{The $\sigma$ symbol represents the standard deviation.}  confidence level contours in three $T_{\rm
  K}$--$n($H$_{2}$) planes centred on the best-fit $N$(CO) value,
where dark-grey dots (red dots in the online version) correspond to
individual models. The $\Delta\chi^{2}$ contours for the HCN and
HCO$^{+}$ gas component are presented in Figure~\ref{fig:app_1}.
Additional contours of $\Delta\chi^2=$50, 100, 200 and 300 are used in
these figures only for guidding the eyes further and see better how
the confidence contours are spreading within the model grid.

With additional (higher J-levels) observations of the same species,
probing denser and warmer gas closer to the forming stars, we should
be able to \emph{i)} confirm the results obtained here and \emph{ii)}
understand better the best-fit model parameter differences across
galaxies (see, e.g., \citealt{Baye04, Baye06} who used the
$^{12}$CO(6-5) and $^{12}$CO(7-6) lines in late-type galaxies).

Despite the uniqueness of the best-fit model for each galaxy
(light-grey filled circles in Fig.~\ref{fig:1}, green filled circles
in the online version), there is a range of models included in the
$1\sigma$ confidence level (darkest zones in Fig.~\ref{fig:1}). These
models can all also be considered as good models, and they could be
used to define formal uncertainties, although these would be
artificially large along the three axes simultaneously and would not
convey the complex shape of the $\Delta\chi^2$ contours. Indeed, as
seen in Figure~\ref{fig:1}, the contours generally have a banana-like
shape similar to what is often obtained in disc galaxies (e.g.\
\citealt{Baye06}), even if the contours are often more ragged
(explained by the fact that there are fewer contraints in our
study). This means that the kinetic temperature and gas volume density
solutions are degenerate. Despite this, we note that the low $\sigma$
confidence level contours are relatively narrow in the 3D parameter
space for IC0676, NGC3032, NGC4526, NGC5866 and UGC09519. The
uncertainties on the best-fit parameters ($T_{\rm K}$, $n$(H$_{2}$),
$N$(CO)) are thus relatively small and the best-fit models identified
are more reliable for these sources.

\begin{table}
  \caption{Best-fit model parameters of the CO gas component.}
  \label{tab:2}
  \begin{tabular}{lrrrr}
    \hline
    Galaxy & $\chi^{2}$ & $T_{\rm K}$ & $n$(H$_{2}$) & $N$(CO) \\
           &            & (K)        & (cm$^{-3}$)  & (cm$^{-2}$) \\
    \hline
    IC0676$^{\rm a}$    &  1.07 &  70 & $10^{3\phantom{.0}}$ & $10^{19.5}$ \\
    IC1024             &  0.56 &  20 & $10^{4\phantom{.0}}$ & $10^{19\phantom{.0}}$ \\
    NGC1222            &  2.06 &  20 & $10^{6.5}$          & $10^{19\phantom{.0}}$ \\
    NGC1266            &  0.85 &  20 & $10^{4\phantom{.0}}$ & $10^{18.5}$ \\
    NGC2764            &  0.11 &  10 & $10^{3.5}$          & $10^{19\phantom{.0}}$\\
    NGC3032            &  5.32 &  10 & $10^{3.5}$          & $10^{19\phantom{.0}}$\\
    NGC3607            &  0.23 &  50 & $10^{3\phantom{.0}}$ & $10^{20\phantom{.0}}$ \\
    NGC3665$^{\rm a}$   &  6.45 &  10 & $10^{3.5}$          & $10^{20\phantom{.0}}$ \\
    NGC4150            &  0.02 &  20 & $10^{3.5}$          & $10^{19\phantom{.0}}$ \\
    NGC4459            &  0.57 &  10 & $10^{4\phantom{.0}}$ & $10^{19.5}$ \\
    NGC4526$^{\rm a}$   &  1.28 &  20 & $10^{3.5}$          & $10^{20\phantom{.0}}$ \\
    NGC4694            &  2.15 &  10 & $10^{3.5}$          & $10^{18.5}$ \\
    NGC4710            &  1.38 &  20 & $10^{3.5}$          & $10^{19.5}$ \\
    NGC5866$^{\rm a}$   & 41.18 &  10 & $10^{7\phantom{.0}}$ & $10^{19.5}$ \\
    NGC6014$^{\rm a}$   &  1.45 & 120 & $10^{3\phantom{.0}}$ & $10^{20\phantom{.0}}$ \\
    NGC7465            &  9.95 & 150 & $10^{4.5}$          &  $10^{20\phantom{.0}}$ \\
    PGC058114$^{\rm a}$ &  0.34 &  30 & $10^{3.5}$          & $10^{19\phantom{.0}}$ \\
    UGC09519$^{\rm a}$  & 25.21 &  40 & $10^{3\phantom{.0}}$ & $10^{19\phantom{.0}}$ \\
    \hline
  \end{tabular}
  
  {\bf Notes:} $^{\rm a}$: Galaxies with a $^{12}$CO(3-2) line
  flux used in the modelling work. 
\end{table}

\begin{figure*}
  \begin{minipage}[c]{.40\linewidth}
    \includegraphics[scale=1.01, clip, trim=0cm 0.5cm 0cm 0cm]{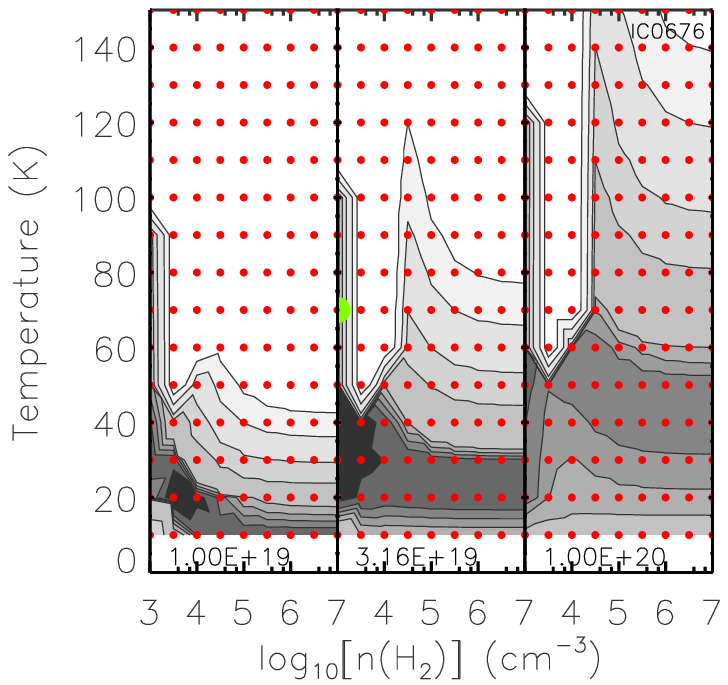}
  \end{minipage}
  \vspace*{-0.6cm}\hspace*{0.1\linewidth}
  \begin{minipage}[c]{.40\linewidth}
    \includegraphics[scale=1.01, clip, trim=0.9cm 0.5cm 0cm 0cm]{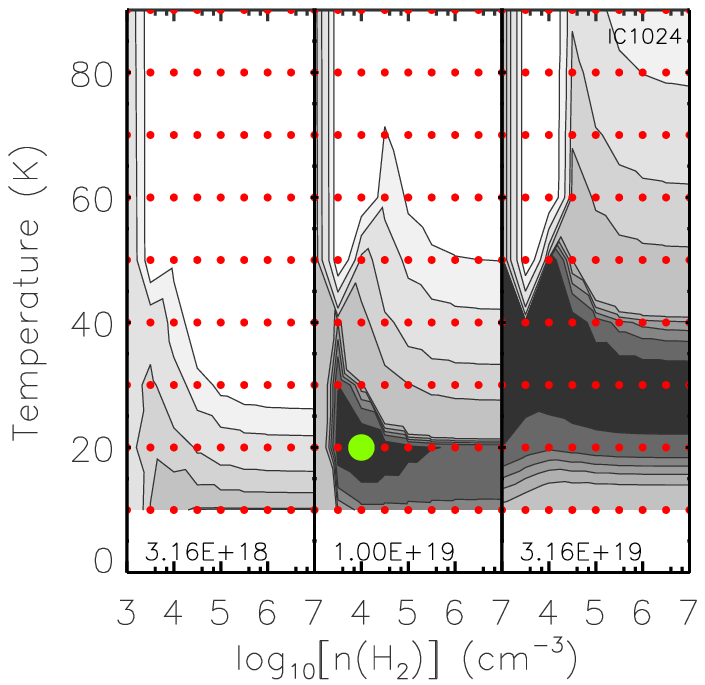}
  \end{minipage}
  \begin{minipage}[c]{.40\linewidth}
    \includegraphics[scale=1.01, clip, trim=0cm 0.5cm 0cm 0cm]{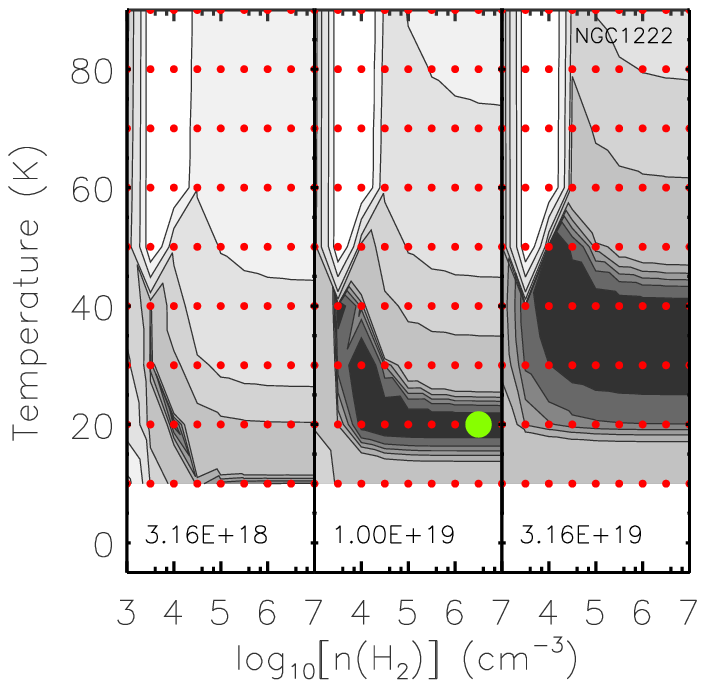}
  \end{minipage}
  \vspace*{-0.6cm}\hspace*{0.1\linewidth}
  \begin{minipage}[c]{.40\linewidth}
    \includegraphics[scale=1.01, clip, trim=0.9cm 0.5cm 0cm 0cm]{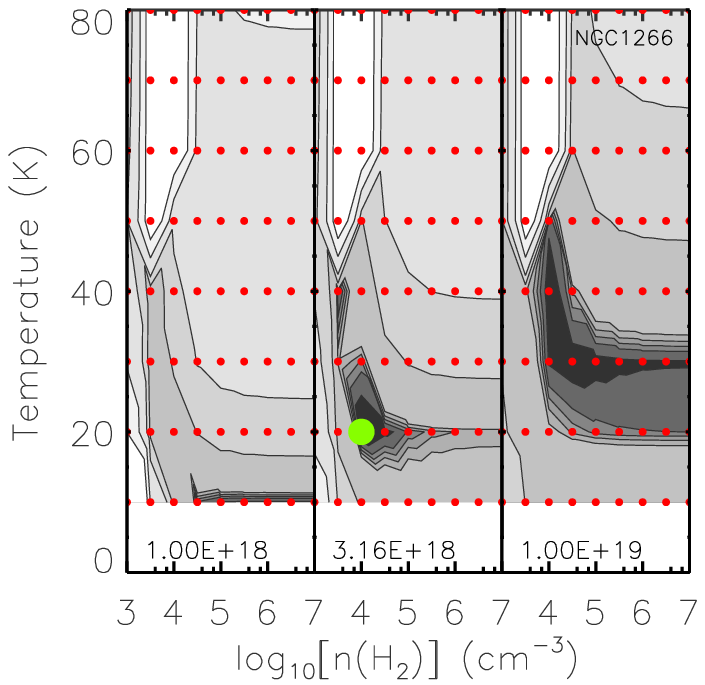}
  \end{minipage}
  \begin{minipage}[c]{.40\linewidth}
    \includegraphics[scale=1.01, clip, trim=0cm 0cm 0cm 0cm]{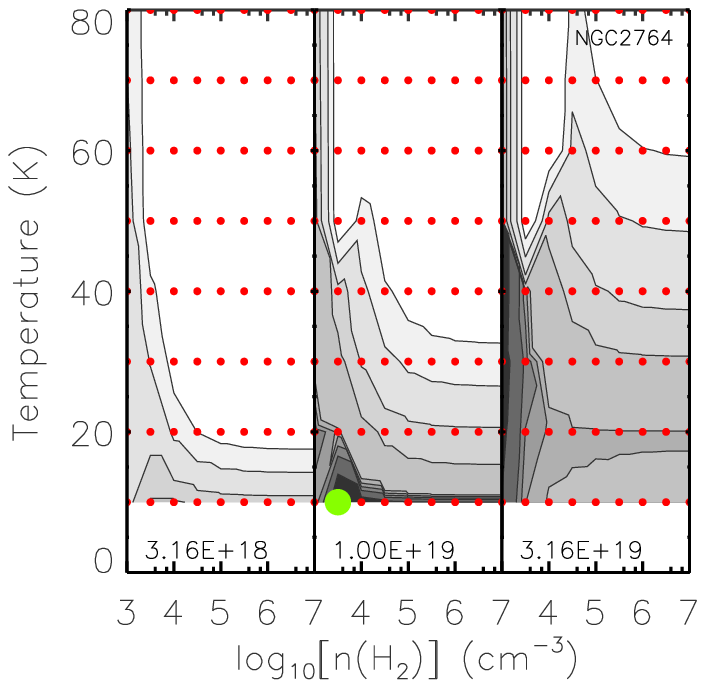}
  \end{minipage}
  \hspace*{0.1\linewidth}
  \begin{minipage}[c]{.40\linewidth}
    \includegraphics[scale=1.01, clip, trim=0.9cm 0cm 0cm 0cm]{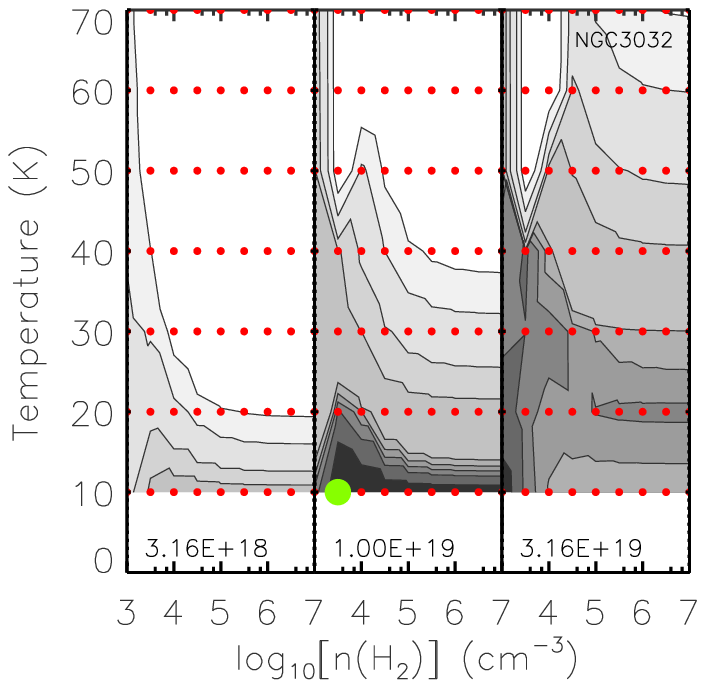}
  \end{minipage}
  \caption{Best-fit models for the CO gas component. Each panel shows
    the $\Delta\chi^2\equiv\chi^2-\chi^2_{\rm min}$ contours of the CO
    gas component as a function of the kinetic temperature $T_{\rm K}$
    and H$_{2}$ volume density $n$(H$_{2}$), for three values of the
    CO column density $N$(CO) (hence three plots) centered on the
    best-fit value and indicated at the bottom of each plot. The model
    grid is shown with dark-grey dots (red dots in the online
    version), whereas the best-fit model of each galaxy
    ($\chi^{2}_{\rm min}$ or $\Delta\chi^{2}=0$) is labeled with a
    light-grey filled circle (green filled circle in the online
    version). The $\Delta\chi^2$ contours and greyscales are for $1$
    to $5\sigma$ confidence levels (i.e.\ $\Delta\chi^2=3.54$, $6.25$,
    $7.81$, $9.35$ and $11.34$ when three line ratios are used and
    $\Delta\chi^2=4.73$, $7.78$, $9.49$, $11.14$ and $13.27$ when four
    line ratios are used). Additional contours of $\Delta\chi^2=$50,
    100, 200 and 300 are used in these
    figures only for guidding the eyes further and see better how the
    confidence contours are spreading within the model grid. The low $\sigma$, high confidence level
    contours containing 'good' models are represented by the darkest
    areas. The galaxy name is indicated in the top-right corner of
    each panel.}
  \label{fig:1}
\end{figure*}

\begin{figure*}
  \addtocounter{figure}{-1}
  \begin{minipage}[c]{.40\linewidth}
    \includegraphics[scale=1.01, clip, trim=0cm 0.5cm 0cm 0cm]{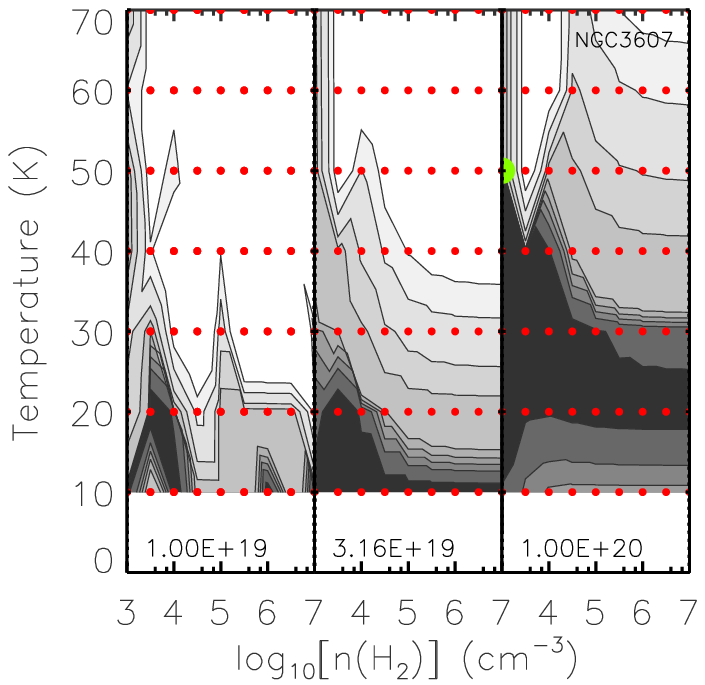}
  \end{minipage}
  \vspace*{-0.6cm}\hspace*{0.1\linewidth}
  \begin{minipage}[c]{.40\linewidth}
    \includegraphics[scale=1.01, clip, trim=0.9cm 0.5cm 0cm 0cm]{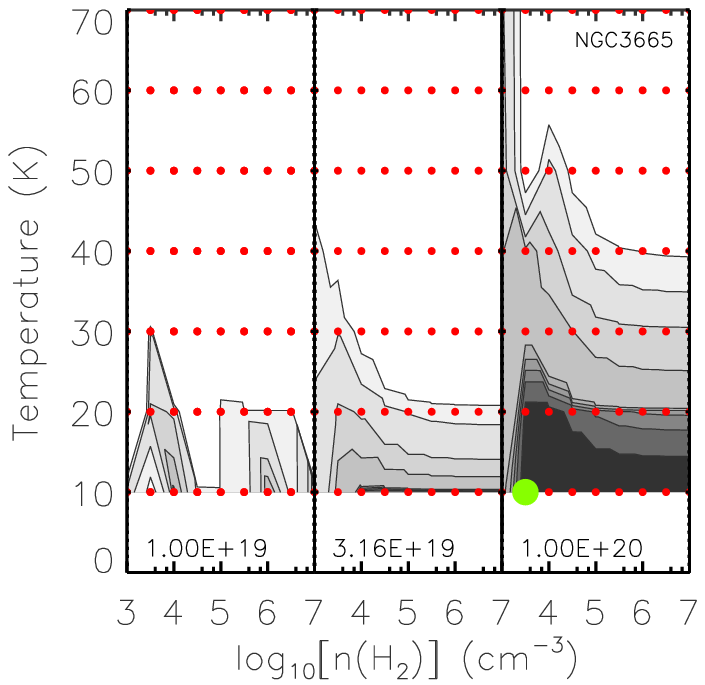}
  \end{minipage}
  \begin{minipage}[c]{.40\linewidth}
    \includegraphics[scale=1.01, clip, trim=0cm 0.5cm 0cm 0cm]{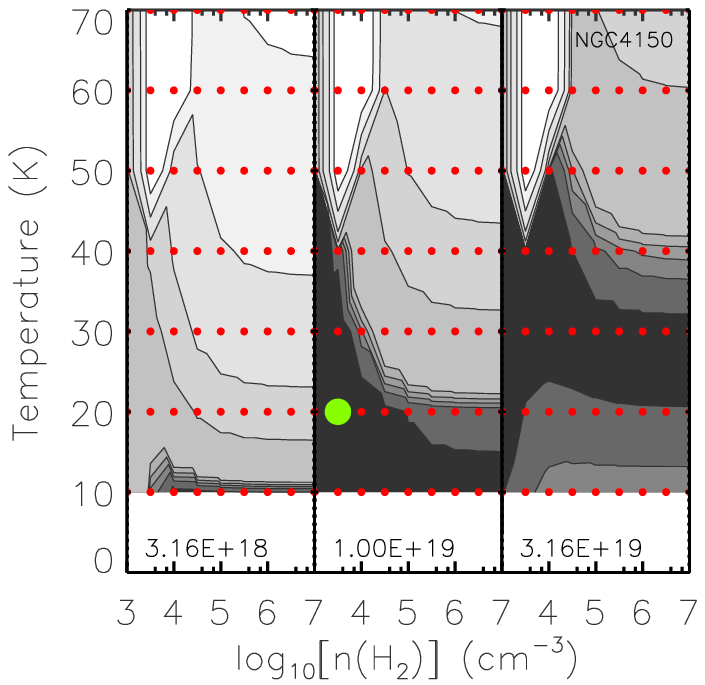}
  \end{minipage}
  \vspace*{-0.6cm}\hspace*{0.1\linewidth}
  \begin{minipage}[c]{.40\linewidth}
    \includegraphics[scale=1.01, clip, trim=0.9cm 0.5cm 0cm 0cm]{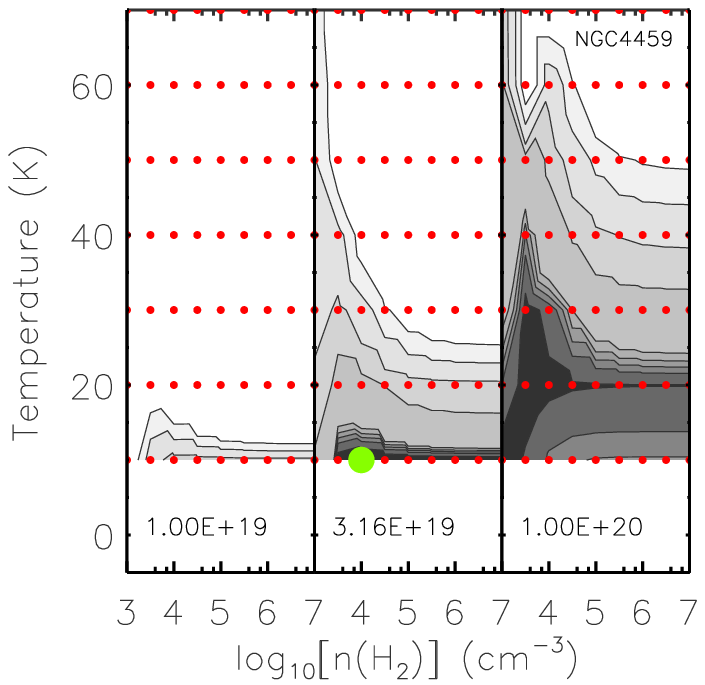}
  \end{minipage}
  \begin{minipage}[c]{.40\linewidth}
    \includegraphics[scale=1.01, clip, trim=0cm 0cm 0cm 0cm]{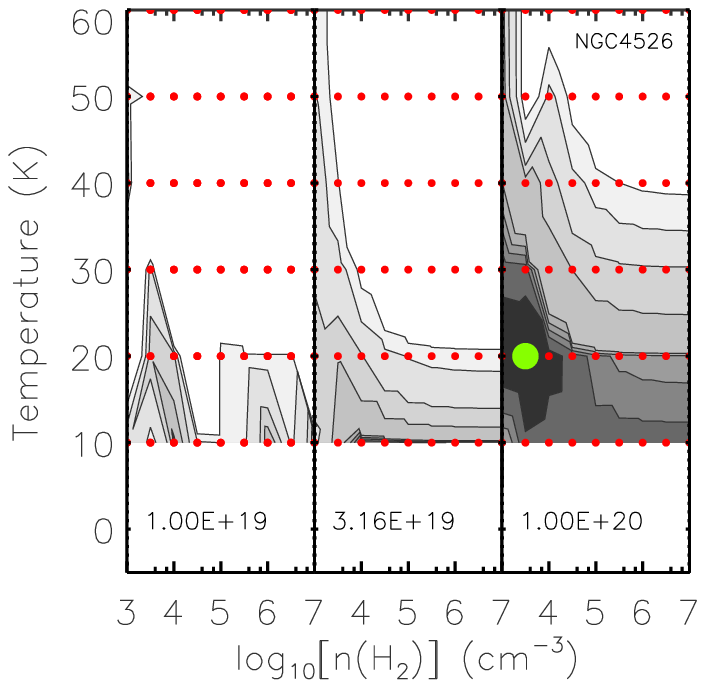}
  \end{minipage}
  \hspace*{0.1\linewidth}
  \begin{minipage}[c]{.40\linewidth}
    \includegraphics[scale=1.01, clip, trim=0.9cm 0cm 0cm 0cm]{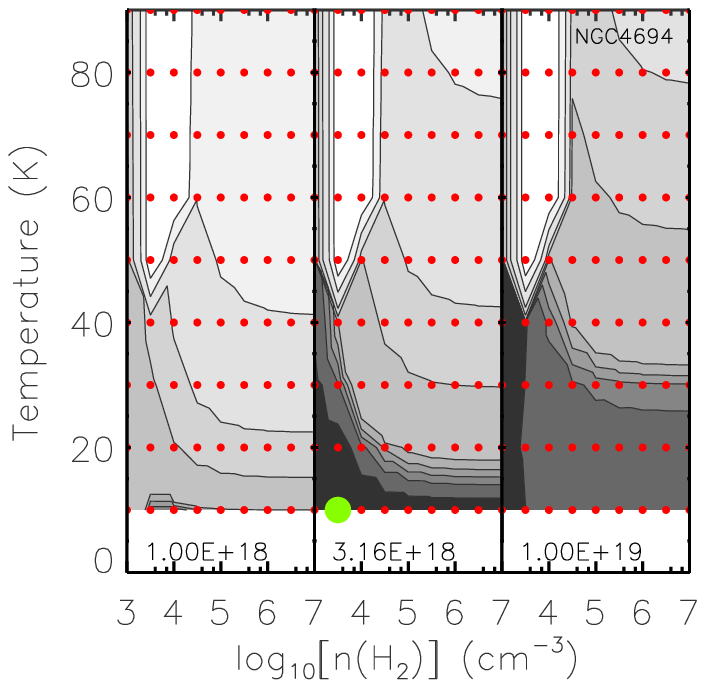}
  \end{minipage}
  \caption{(Continued).}
\end{figure*}

\begin{figure*}
  \addtocounter{figure}{-1}
  \begin{minipage}[c]{.40\linewidth}
    \includegraphics[scale=1.01, clip, trim=0cm 0.5cm 0cm 0cm]{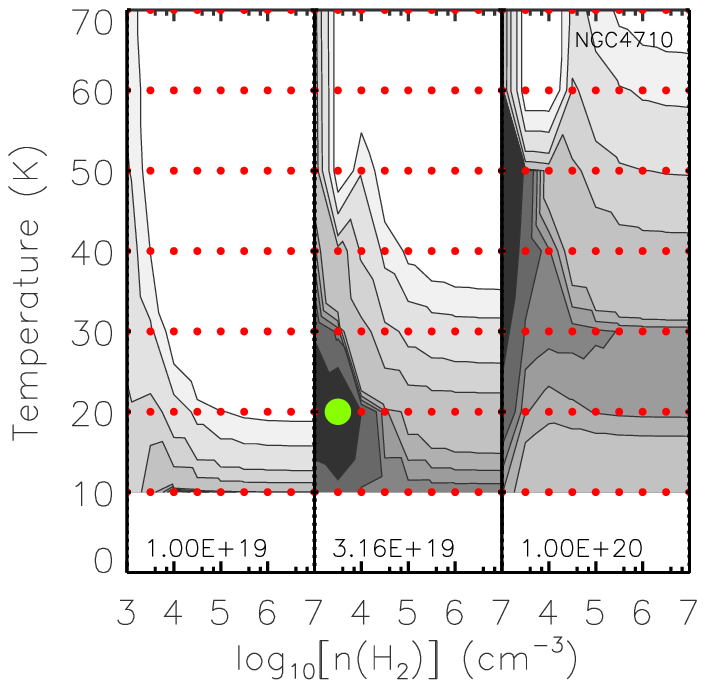}
  \end{minipage}
  \vspace*{-0.6cm}\hspace*{0.1\linewidth}
  \begin{minipage}[c]{.40\linewidth}
    \includegraphics[scale=1.01, clip, trim=0.9cm 0.5cm 0cm 0cm]{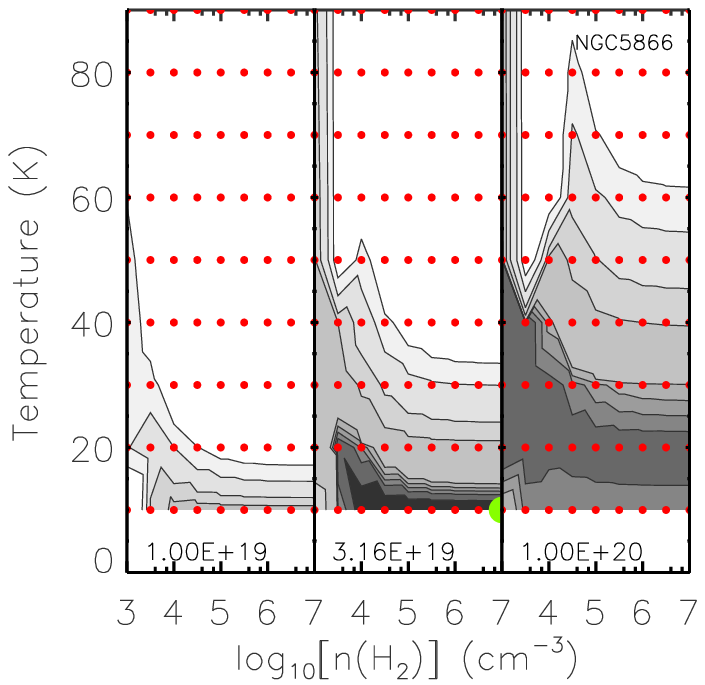}
  \end{minipage}
  \begin{minipage}[c]{.40\linewidth}
    \includegraphics[scale=1.01, clip, trim=0cm 0.5cm 0cm 0cm]{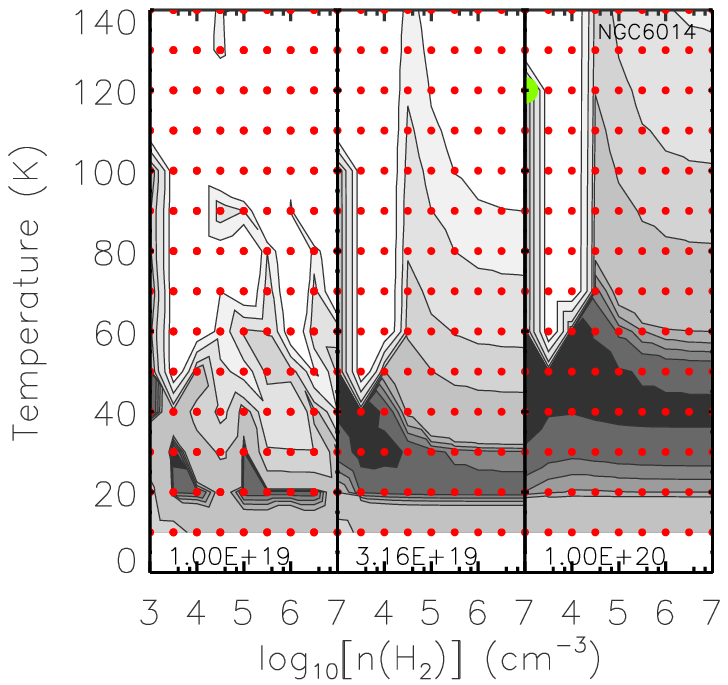}
  \end{minipage}
  \vspace*{-0.6cm}\hspace*{0.1\linewidth}
  \begin{minipage}[c]{.40\linewidth}
    \includegraphics[scale=1.01, clip, trim=0.9cm 0.5cm 0cm 0cm]{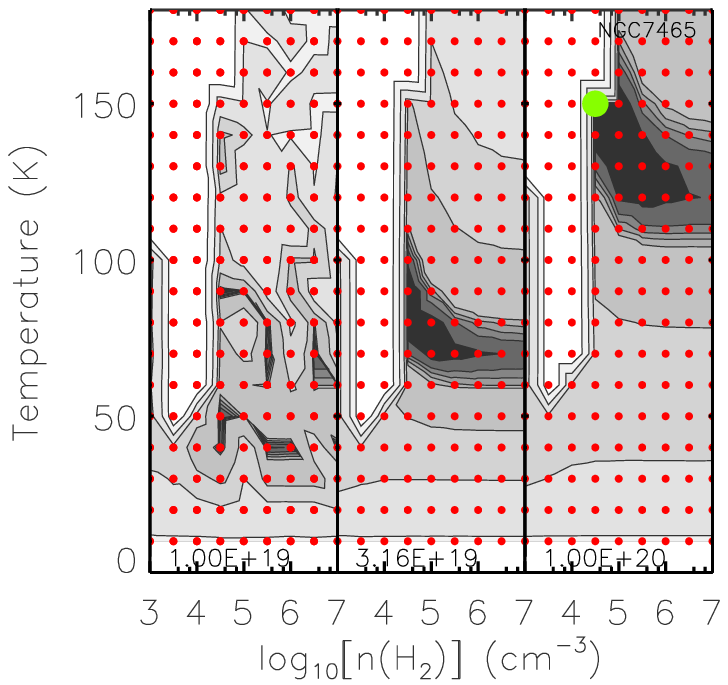}
  \end{minipage}
  \begin{minipage}[c]{.40\linewidth}
    \includegraphics[scale=1.01, clip, trim=0cm 0cm 0cm 0cm]{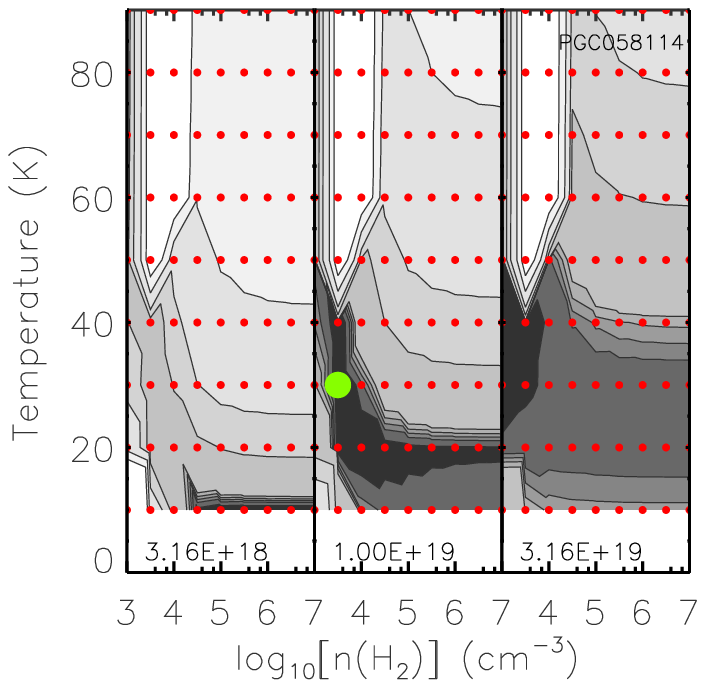}
  \end{minipage}
  \hspace*{0.1\linewidth}
  \begin{minipage}[c]{.40\linewidth}
    \includegraphics[scale=1.01, clip, trim=0.9cm 0cm 0cm 0cm]{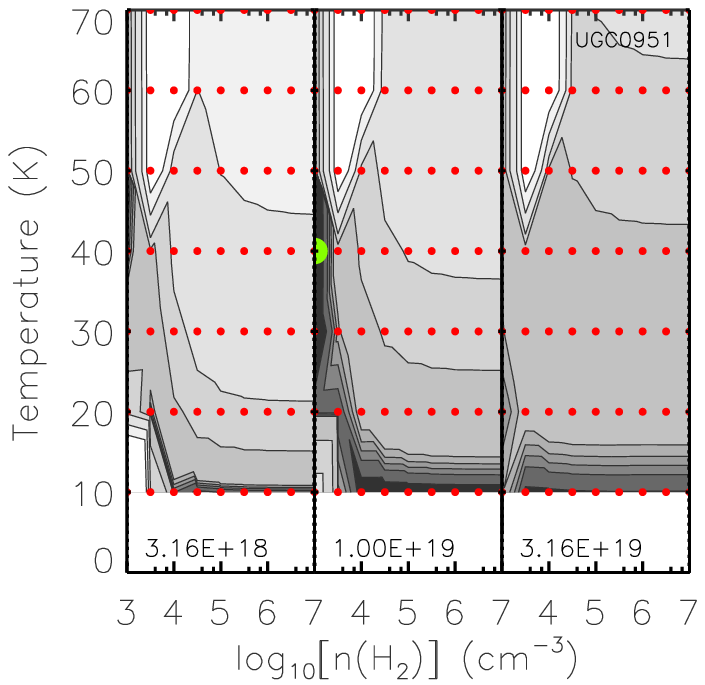}
  \end{minipage}
  \caption{(Continued).}
\end{figure*}

\subsection{Most likely model identification}\label{subsec:likely}

To better understand the distribution of good models in our 3D
parameter space, and to define robust uncertainties, we have
investigated a complementary 'best' model identification method that
consists in determining the marginalised probability distribution
functions (PDFs) of the three input model parameters ($T_{\rm K}$,
$n$(H$_{2}$) and $N$(CO)), following the formalism developed in
\citet{Kavi07}. For each parameter value, we have calculated the
likelihood, i.e.\ the sum of the exponentials of $-\Delta\chi^2/2$,
marginalizing over the other two free parameters. This technique
allows us to isolate the effect of each parameter, and the results are
shown in Figure~\ref{fig:4}.

From the plots shown in Figure~\ref{fig:4}, we have calculated the
total area defined by each PDF, terminating the integration when
necessary at the end of our model grid. This area represents the total
probability ($100\%$) of each parameter, from which we have determined
the mean. This defines the equipartition ($50\%$-$50\%$) of the PDF
and, for well-behaved (i.e. single-peaked and symmetric) PDFs, is usually considered the most natural way
to identify the most likely model. The mean is shown in each plot of
Figure~\ref{fig:4} by a solid grey line surrounded by two dashed grey
lines, themselves defining the $1\sigma$ uncertainty ($68\%$
probability, $34\%$ on each side of the mean). The mean and associated
$1\sigma$ uncertainties are listed in Table~\ref{tab:3} for each model
parameter and galaxy.

Similarly to the best-fit method, the likelihood method yields groups
of galaxies with similar ranges of kinetic temperatures. These
categories are however less marked than for the best-fit models (see
Section~\ref{subsec:chi} and Table~\ref{tab:2}). The H$_{2}$ volume
densities and CO column densities are again rather uniform (variations
of less than a factor of $10$ for most sources, excluding NGC1222 and
NGC5866), similar to what is observed with the best-fit method.

\begin{table}
  \caption{Mean likelihood parameters and their $1\sigma$
    uncertainties.}
  \label{tab:3}
  \begin{tabular}{lrrr}
    \hline
    Galaxy & $T_{\rm K}$ & $n$(H$_{2}$) & $N$(CO) \\
           & (K)        & (cm$^{-3}$)  & (cm$^{-2}$) \\
    \hline
    IC0676$^{\rm a}$  & $55.8_{-32.1}^{+34.1}$  & $1.9^{\phantom{0}+23.4}_{\phantom{00}-0.8}\times 10^{3}$ & $3.0^{+2.1}_{-1.5}\times 10^{19}$ \\
    IC1024$^{\rm a}$  & $39.6_{-15.5}^{+27.7}$  & $3.6^{+147.0}_{\phantom{00}-3.4}\times 10^{4}$ & $3.1^{+3.2}_{-1.8}\times 10^{19}$ \\
    NGC1222$^{\rm a}$ & $34.8_{-14.9}^{+25.4}$  & $2.3^{\phantom{0}+27.9}_{\phantom{00}-2.1}\times 10^{5}$ & $2.1^{+4.0}_{-1.3}\times 10^{19}$ \\
    NGC1266$^{\rm a}$ & $34.7_{-15.7}^{+36.1}$  & $8.1^{+211.8}_{\phantom{00}-7.1}\times 10^{4}$ & $8.7^{+3.0}_{-6.7}\times 10^{18}$ \\
    NGC2764          & $18.3_{\phantom{0}-6.5}^{+39.8}$  & $4.0^{+112.6}_{\phantom{00}-2.5}\times 10^{3}$ & $1.4^{+2.0}_{-0.7}\times 10^{19}$ \\
    NGC3032          & $13.7_{\phantom{0}-2.6}^{\phantom{0}+5.1}$  & $2.1^{+132.3}_{\phantom{00}-1.9}\times 10^{4}$ & $1.1^{+1.0}_{-0.5}\times 10^{19}$ \\
    NGC3607          & $33.8_{-17.1}^{+29.2}$  & $4.4^{+397.8}_{\phantom{00}-3.0}\times 10^{3}$& $5.5^{+3.0}_{-2.7}\times 10^{19}$ \\
    NGC3665          & $13.4_{\phantom{0}-2.4}^{\phantom{0}+3.8}$  & $7.2^{+195.4}_{\phantom{00}-6.7}\times 10^{4}$ & $7.0^{+2.0}_{-2.2}\times 10^{19}$ \\
    NGC4150$^{\rm a}$ & $43.6_{-23.7}^{+41.5}$  & $6.3^{+563.3}_{\phantom{00}-4.9}\times 10^{3}$& $2.5^{+3.1}_{-1.6}\times 10^{19}$ \\
    NGC4459          & $14.6_{\phantom{0}-3.4}^{\phantom{0}+6.8}$ & $5.9^{+180.8}_{\phantom{00}-5.4}\times 10^{4}$ & $3.5^{+2.8}_{-1.5}\times 10^{19}$ \\
    NGC4526          & $19.5_{\phantom{0}-5.1}^{\phantom{0}+5.0}$  & $3.9^{\phantom{0}+25.4}_{\phantom{00}-2.2}\times 10^{3}$ & $6.7^{+2.3}_{-2.5}\times 10^{19}$ \\
    NGC4694          & $23.1_{-10.6}^{+45.9}$   & $8.7^{+951.2}_{\phantom{00}-7.2}\times 10^{3}$ & $5.7^{+1.3}_{-3.2}\times 10^{18}$ \\
    NGC4710$^{\rm a}$ & $51.3_{-32.0}^{+37.0}$  & $1.9^{\phantom{00}+3.0}_{\phantom{00}-0.7}\times 10^{3}$ & $5.1^{+3.2}_{-2.6}\times 10^{19}$ \\
    NGC5866          & $14.0_{\phantom{0}-2.8}^{\phantom{0}+6.5}$    & $2.9^{\phantom{0}+29.6}_{\phantom{00}-2.7}\times 10^{5}$ & $3.3^{+2.4}_{-1.3}\times 10^{19}$ \\
    NGC6014$^{\rm a}$ & $65.1_{-27.7}^{+42.0}$  & $2.5^{+135.0}_{\phantom{00}-1.2}\times 10^{3}$ & $6.0^{+2.7}_{-2.9}\times 10^{19}$ \\
    NGC7465$^{a}$    & $124.3_{-55.0}^{+21.5}$  & $8.3^{\phantom{0}+38.6}_{\phantom{00}-5.5}\times 10^{4}$ & $4.3^{+3.6}_{-2.9}\times 10^{19}$ \\
    PGC058114       & $35.0_{-19.1}^{+48.0}$& $1.2^{+143.3}_{\phantom{00}-1.0}\times 10^{4}$ & $1.4^{+2.7}_{-1.1}\times 10^{19}$ \\
    UGC09519        & $33.9_{-21.4}^{+23.2}$   & $2.3^{+119.1}_{\phantom{00}-2.2}\times 10^{4}$ & $1.1^{+1.3}_{-0.5}\times 10^{19}$ \\
    \hline
  \end{tabular}
  
  {\bf Notes:} $^{a}$: Sources with a marginalised probability distribution
  function with two or more peaks, where the mean parameter listed in
  the table is misleading, i.e.\ it is not the most likely parameter
  (see text in Section~\ref{subsec:likely}).
\end{table}

The likelihood method helps us to identify the `best' model in a
complementary way to the simple $\chi^{2}$ method. Comparing
Tables~\ref{tab:2} and \ref{tab:3} and the positions of the black dots
(best-fit models) and solid grey lines (most likely models) in
Figure~\ref{fig:4}, we see that in the majority of sources the two
methods yield consistent results within the (admittedly rather large)
uncertainties. Indeed, one notes that the highest PDF peaks agree well
with the best-fit model paramaters. In a few cases (NGC1266, NGC6014
and NGC7465 are the best examples), however, the most likely models do
not exactly correspond to the best-fit models. The apparent
discrepancy between the best-fit and most likely models is then
generally due to the double-peaked (or triple-peaked for NGC7465)
shape of the kinetic temperature PDF (see the left panels in
Fig.~\ref{fig:4}). These two (three) peaks, when of almost equal
amplitudes, imply that there are two (three) almost equally probable
solutions. The derived mean for these distributions, located at a
roughly equal distance from the two (three) peaks, is therefore
misleading and does not generally correspond to the most likely model.

\begin{figure*}
    \centering
    \includegraphics[scale=0.8]{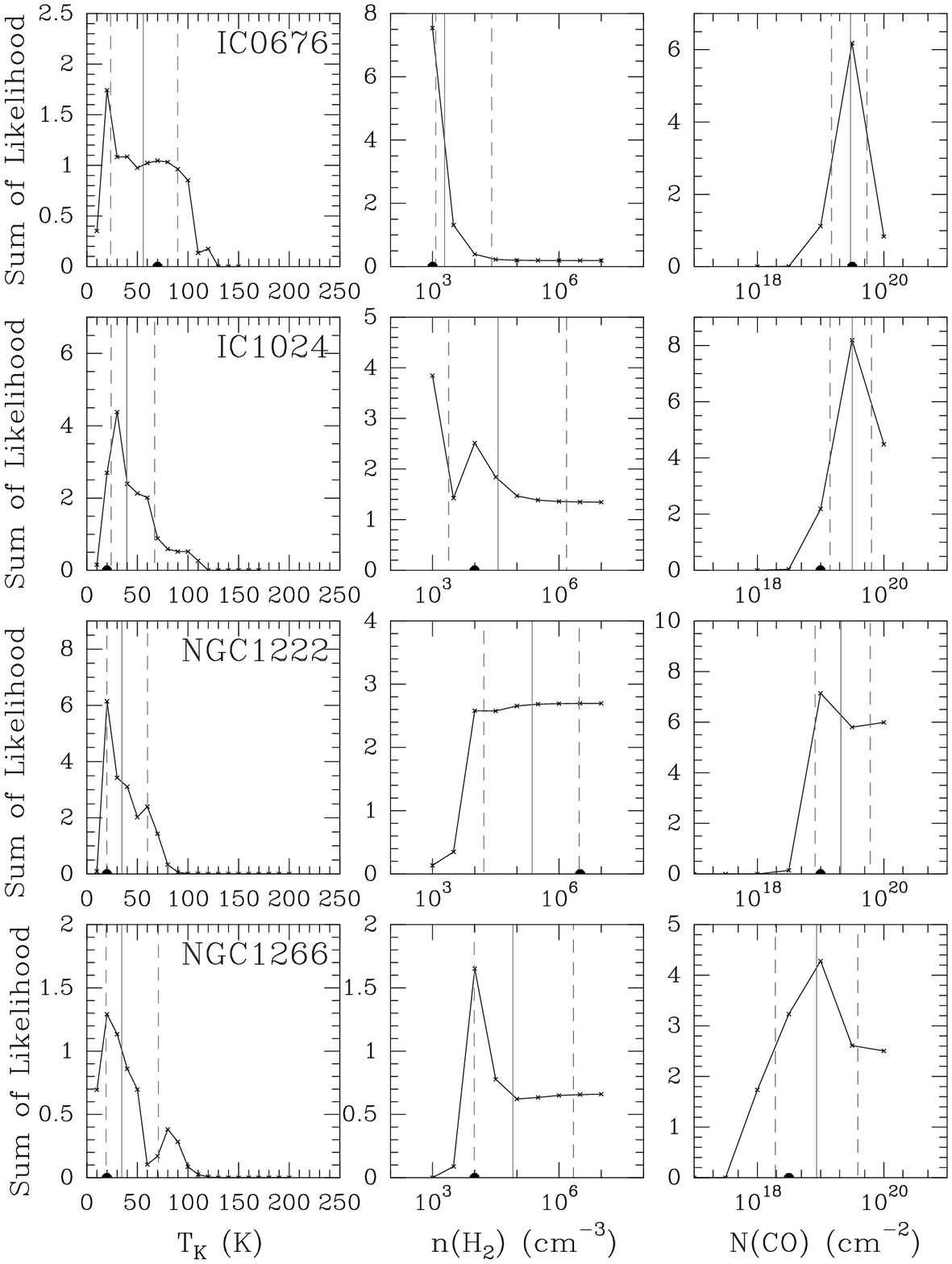}
    \caption{Marginalised probability distribution functions of our
      model parameters. For each galaxy (indicated in the top-right
      corner of the left-hand plots), the PDF of each model parameter
      is shown, marginalised over the two others. The best-fit
      ($\chi^{2}_{\rm min}$) model parameters are indicated by solid
      black circles on the $x$ axes. The mean likelihood values and
      $1\sigma$ uncertainties are represented by solid and dashed grey
      lines, respectively.}
    \label{fig:4}
\end{figure*}

\begin{figure*}
    \addtocounter{figure}{-1}
    \centering
    \includegraphics[scale=0.8]{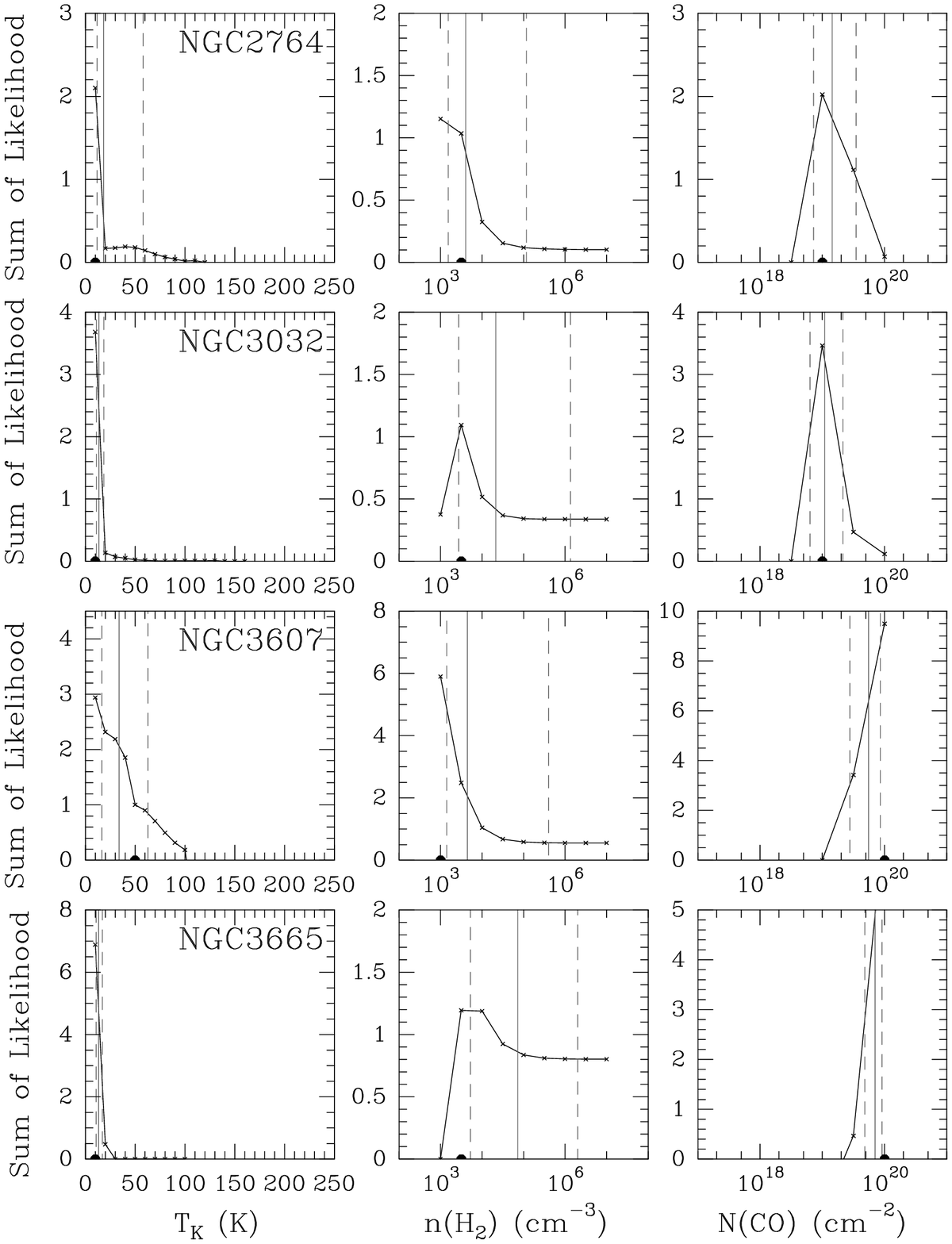}
    \caption{(Continued).}\label{fig:5}
\end{figure*}

\begin{figure*}
    \addtocounter{figure}{-1}
    \centering
    \includegraphics[scale=0.8]{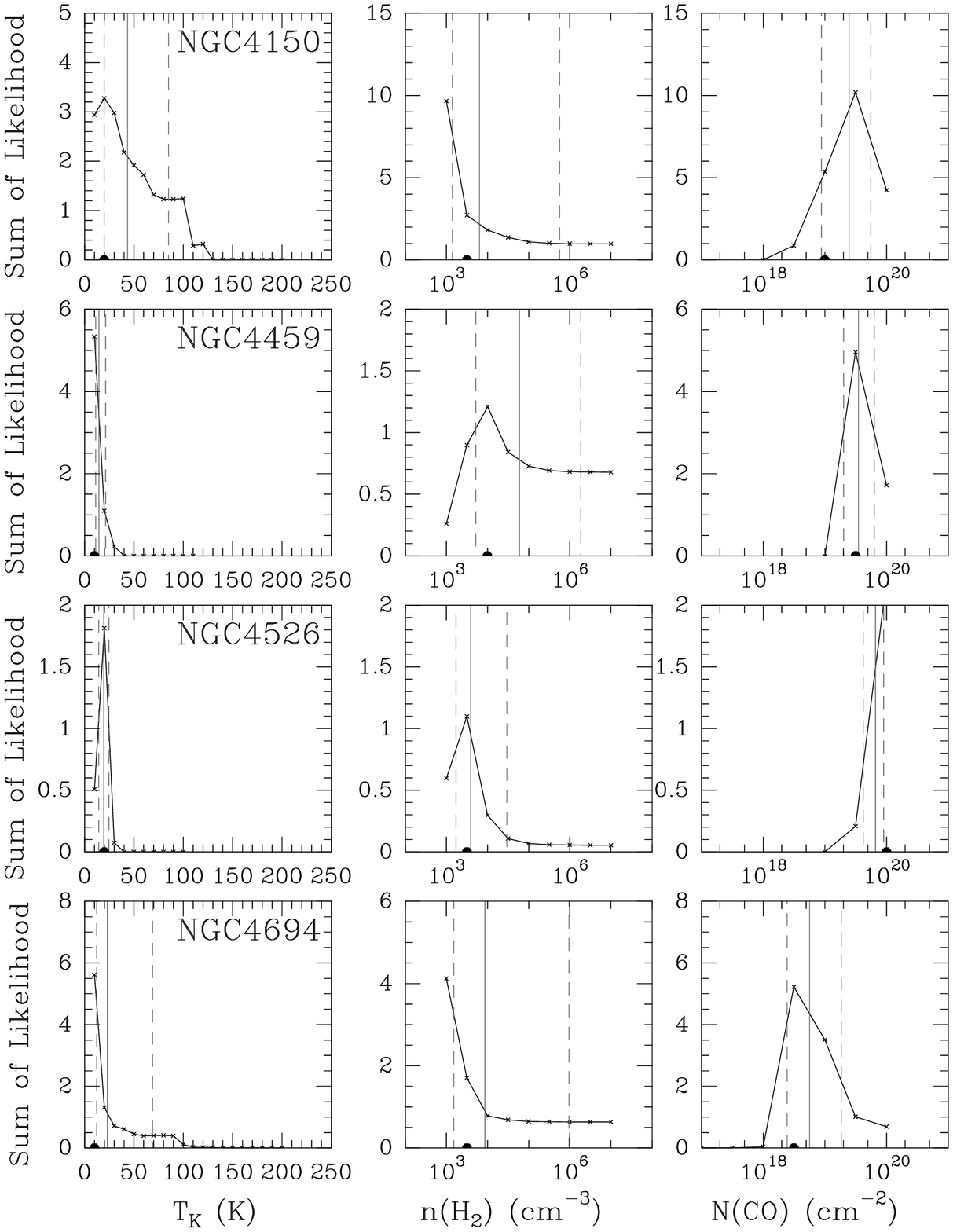}
    \caption{(Continued).}\label{fig:6}
\end{figure*}

\begin{figure*}
    \addtocounter{figure}{-1}
    \centering
    \includegraphics[scale=0.8]{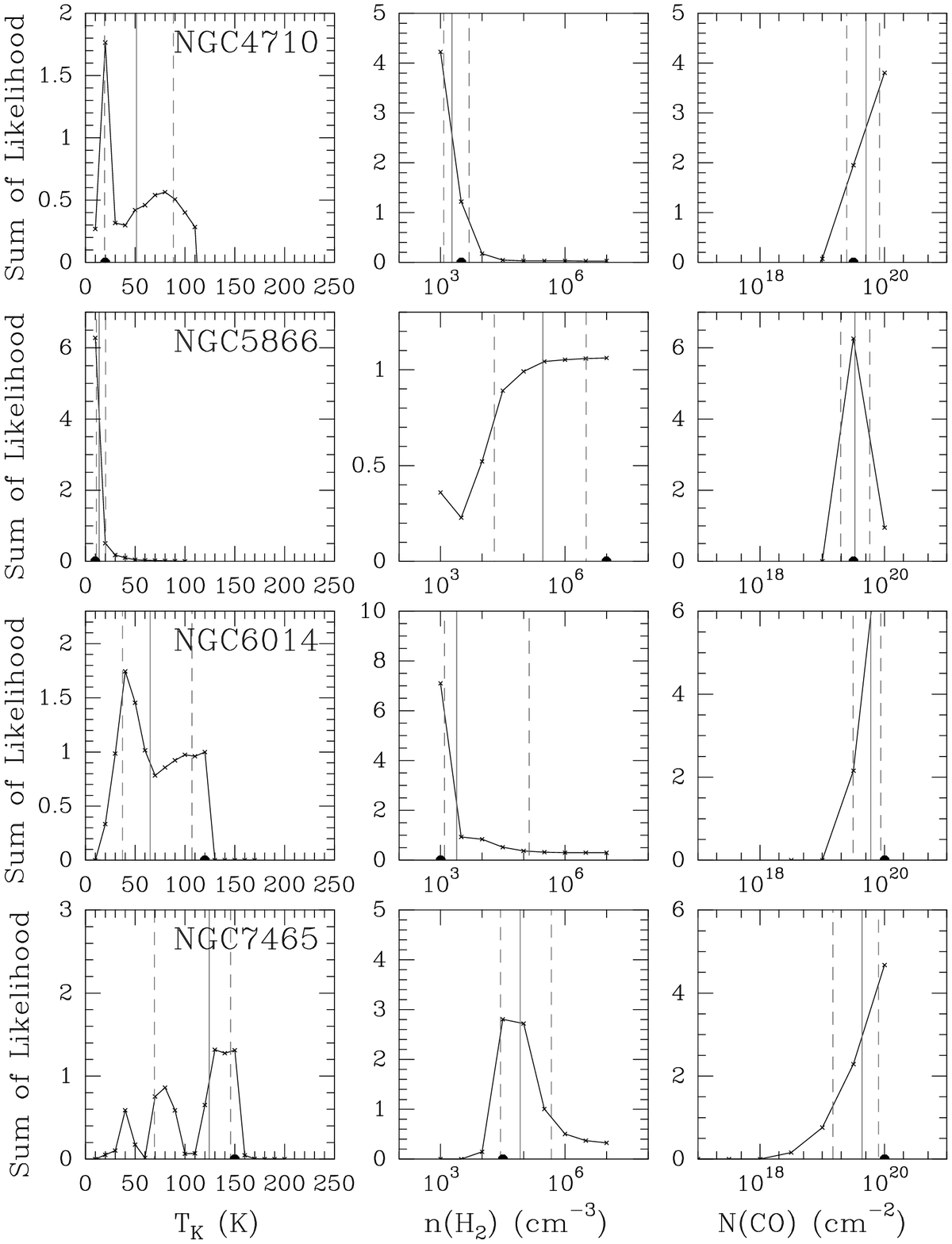}
    \caption{(Continued).}\label{fig:7}
\end{figure*}

\begin{figure*}
    \addtocounter{figure}{-1}
    \centering
    \includegraphics[scale=0.8,clip, trim=0cm 0.68cm 0cm 0cm]{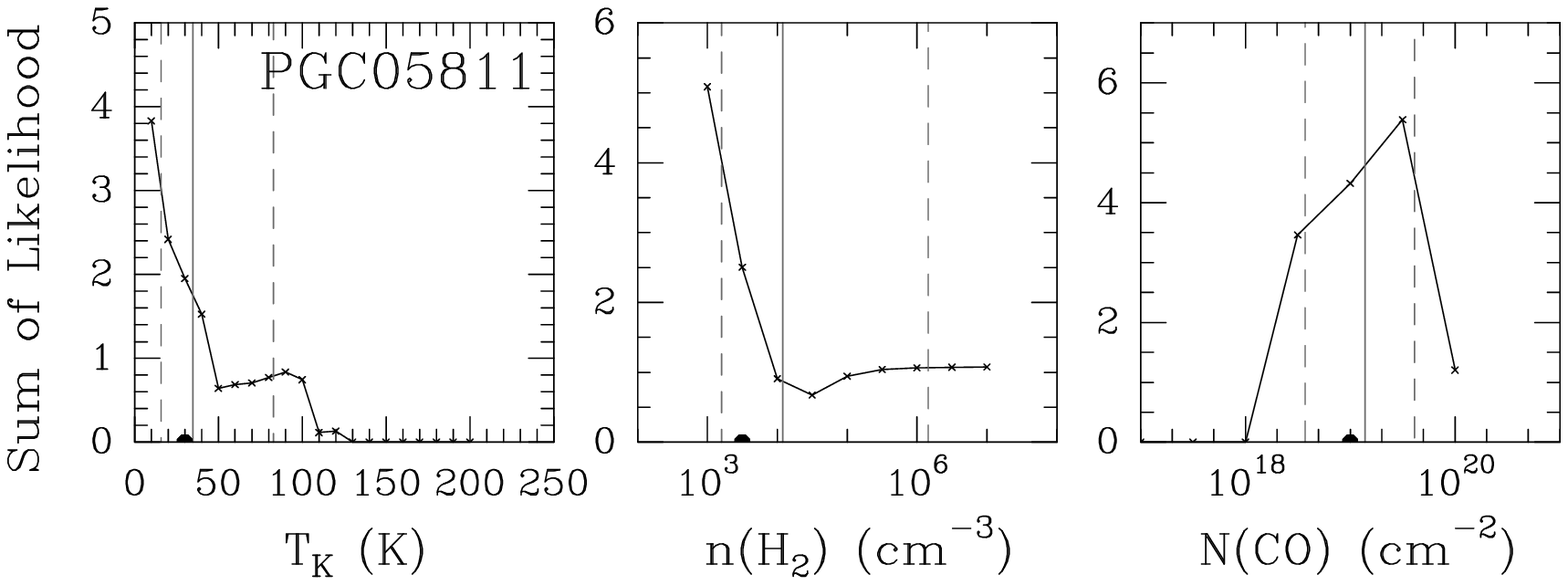}
    \includegraphics[scale=0.8]{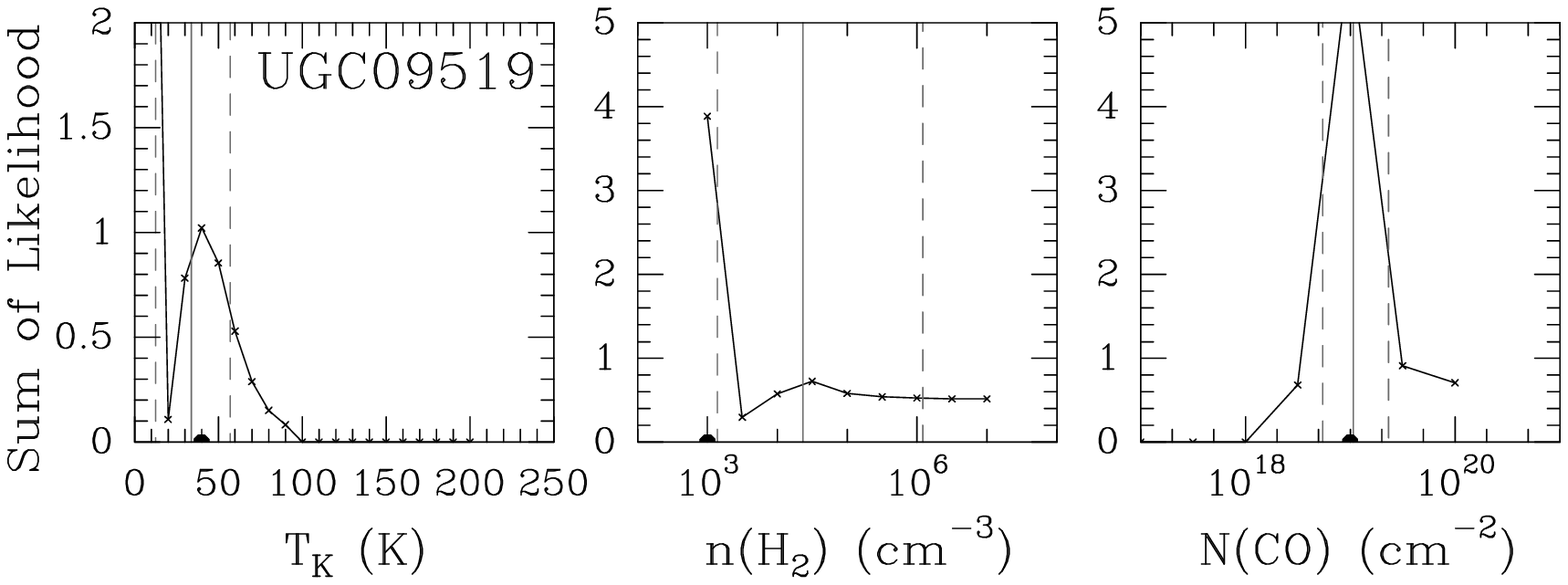}
    \caption{(Continued).}\label{fig:8}
\end{figure*}

\subsection{ETG CO spectral line energy distributions}\label{subsec:SED}

Calculating models up to the $15^{\rm th}$ level population for the CO
molecule allows us to predict velocity-integrated line intensities up
to $^{12}$CO(15-14). Converting the velocity-integrated line
intensities (in K~km~s$^{-1}$) into intensities (in
W~m$^{-2}$~sr$^{-1}$) using Eq.~4 in \citet{Baye04}, and plotting both
quantities as a function of the higher rotational level of the
transition considered (hereafter J$_{\rm upper}$), yields the
predicted CO SLED of our ETGs (see Fig.~\ref{fig:9}).

In Figure~\ref{fig:9}, we show the best-fit model SLED as a solid
black line and use two solid grey lines to delineate the range of
SLEDs associated with a $1\sigma$ confidence level on the model
parameters. In other words, all the models included within the first
$\Delta\chi^{2}$ contour in Figure~\ref{fig:1} have their SLEDs
between the two solid grey lines. The SLED corresponding to the most
likely model is also shown as a dotted black line. Unsurprisingly, the
largest $1\sigma$ SLED ranges occur for sources with a large kinetic
temperature uncertainty (i.e.\ when the 3D darkest region in
Fig.~\ref{fig:1} has a large extent along the $y$ axis). For example,
within the $1\sigma$ confidence level of its model parameters (see
Fig.~\ref{fig:1}), the predicted CO SLED peak position (hereafter
J$_{\rm max}$) of NGC6014 varies between J$_{\rm upper}=5$ and
$7$, covering three increments in J$_{\rm upper}$, while its kinetic
temperature varies by a factor of $6$. A similar behaviour is observed
for NGC1266, NGC4694, NGC4710, NGC7465 and PGC058114, with a predicted
SLED turnover position increment of $2$ or $3$ and a kinetic
temperature increase by a factor of $3$ to $6$ within $1\sigma$. When
$T_{\rm kin}$ does not vary by more than a factor of $3$, J$_{\rm
  max}$ only varies by one increment in J$_{\rm upper}$ or less (e.g.\
IC0676 and IC1024). This provides a clear indication of how the peak
of a galaxy SLED varies with kinematic temperature, and appears
consistent with other current theoretical studies (e.g.\ Wu et al., in
prep.). Of course, the H$_{2}$ volume density also has an important
impact on the shape of the SLEDs, but with our present galaxy sample
the corresponding effect is less clear (see Fig.~\ref{fig:9}). It is
however generally thought that when the gas volume density increases,
the SLED widens (Wu et al.\, in prep.).

To our knowledge, ours are the first CO SLEDs ever produced for normal
ETGs (but see \citealt{Ecka90a}, \citealt{Wild97} and \citealt{Mull09}
for modelling results on Centaurus~A). They are thus of prime
importance to better understand the star formation activity of these
systems. Indeed, J$_{\rm max}$ can be used as a tracer of a galaxy's
star formation activity, as it is this activity that directly drives
the gas excitation in most sources, including those studied
here. As seen in \citet{Weis07} for late-type galaxies, the
  position of the SLED turnover appears linked to the star formation
  rate. For example, the galaxy APM0879
  (SFR$\,\gg100$~$M_{\odot}$~yr$^{-1}$) has the highest observed SLED
  turnover position, whereas the Milky Way
  (SFR$\,\approx1$~$M_{\odot}$~yr$^{-1}$) has the lowest. It has also
been noted that the SLEDs of nearby ($<10$~Mpc) starburst galaxies
peak at higher transitions (around (7-6)) on average than those of
nearby normal spirals (such as the Milky Way, whose SLED turnover is
located at around the (4-3) transition; see, e.g., \citealt{Baye04,
  Baye06}). Nevertheless, there is currently no study showing
  a direct link between a SLED turnover position and the SFR. What
  drives the position of the turnover is thought to be not the
  kinematic temperature alone, but rather a combination of the
  kinematic temperature and gas volume density. The gas column density
  is expected to impact mostly on the intensity of the SLED turnover,
  not its position. The $^{12}$C/$^{13}$C fractional abundance (for
  CO), line width $\Delta v$ and opacity can also influence the SLED
  turnover position, but they are secondary order effects (again, see
  Wu et al.\, in prep.).

We present in Table~\ref{tab:3} a summary of the predicted SLED
turnover positions (J$_{\rm max}$) for our galaxies, derived from the
best-fit models (best-fit and $1\sigma$ uncertainties). We
  also show in Figure~\ref{fig:2} (top) those SLED turnover positions
  as a function of each of the best-fit model parameters ($T_{\rm K}$,
  $n$(H$_{2}$) and $N$(CO)). There are weak trends of J$_{\rm max}$
  with both $T_{\rm K}$ and $N$(CO). Those trends however need to be
  confirmed with a more reliable (i.e.\ larger) statistical sample, as
  they are mostly driven by the modelling results of a single galaxy
  (NGC7465, with the highest $T_{\rm kin}$ and J$_{\rm max}$).  In
addition, with only the (1-0), (2-1) and occasionally the (3-2)
transition, it is often difficult to reliably predict the existence
and/or position of a SLED peak, and therefore to confidently infer the
level of star formation activity. The results presented here, although
suggestive and exciting since available for the first time, are thus
based on limited data and a simple theoretical model with many
limitations and assumptions, and should therefore be considered
accordingly.

Recently, \citet{Shap10} and Falcon-Barroso et al.\ (in prep.)
  derived preliminary star formation rates for $8$ of the molecular
  gas-rich ETGs in our study (NGC1266, NGC3032, NGC3607, NGC4150,
  NGC4459, NGC4526, NGC4694 and NGC5866), with SFRs ranging from
  $0.06$ to $0.59$~$M_{\odot}$~yr$^{-1}$ (see
  Table~\ref{tab:5}). These estimates are based on non-stellar
  $8~\mu$m emission. Obtaining SFR estimates for ETGs is however
  fraught with danger, as most usual SFR indicators are not
  straightforwardly applicable to ETGs. At the very least, specific
  properties of ETGs (e.g.\ UV-upturn population, X-ray halos, etc)
  may significantly alter the calibration factors used in those
  methods (see the discussion in \citealt{Croc11}). Figure~\ref{fig:2}
  (bottom) shows the SLED turnover position as a function of the
  $8$~$\mu$m-derived SFR for those $8$ galaxies. No clear trend is
  observed. Plots showing these SFR estimates as a function of the
  best-fit model parameters ($T_{\rm K}$, $n$(H$_{2}$) and $N$(CO))
  show no clear trend either (Fig.~\ref{fig:3}). The sample size is
  however rather small, preventing us from drawing reliable
  conclusions. Whether or not ETG SLED turnover positions follow the
  same trend with SFR as those of late-type galaxies
  \citep[e.g.][]{Weis07, Baye04, Baye06} thus remains to be
  confirmed. Because of the complicating issues with ETG SFRs
  mentioned above, a detailed study of the SFRs of our sample galaxie
  is beyond the scope of this paper, and we refer the reader to
  \citet{Shap10} and Falcon-Barroso et al.\ (in prep.) for further
  discussions.

Despite their limitations, our models are nevertheless very useful
tools, making specific predictions and paving the way for future
observations. These predictions can in turn inform on the models
themselves. For instance, the two methods of 'best' model
identification yield different predicted turnover positions in some
sources. If the $^{12}$CO(6-5) line was detected in IC1024 and
NGC1266, for example, with flux in both cases below
$10^{-10}$~W~m$^{-2}$~sr$^{-1}$, a flat SLED seems an unlikely
possibility for those two galaxies. In such case, there should rather
be a SLED turnover and it should be located between J=2-1 and J=6-5. On the
contrary, if the $^{12}$CO(6-5) line was detected in these two
galaxies but with a flux of about $10^{-9}$~W~m$^{-2}$~sr$^{-1}$,
this will favour a flat SLED scenario thereby indicating that more sophisticated models
are required to infer the properties of the molecular gas (e.g.\
photon-dominated or cosmic ray-dominated region codes such as those
developed by \citealt{Lepe06,Baye09a,Baye10c,Papa10,Meij11}). Finally,
the great advantage of detecting the $^{12}$CO(6-5) line, in most
sources displayed in Figures ~\ref{fig:9} would be to
single out the most accurate method of best model identification in
ETGs (either $\chi^{2}$ or maximum likelihood).

Further limits to the accuracy of the predicted SLEDs arise from the
underlying assumptions, that have to be carefully considered. For
example, the shapes of the line profiles are neglected when working
with line ratios (see Section~\ref{subsec:param}). However, this is
not true anymore when modelling velocity-integrated line intensities
or intensities, and unfortunately RADEX assumes a Gaussian line
profile \citep{VanderTak07}. We know that for galaxies such as
NGC3032, NGC3607, NGC3665, NGC4459, NGC4526, NGC5866 and NGC6014, a
double-peaked line profile is more realistic, so for these sources the
SLEDs shown in Figure~\ref{fig:9} are more uncertain.

\begin{figure*}
  \centering
  \includegraphics[scale=0.8]{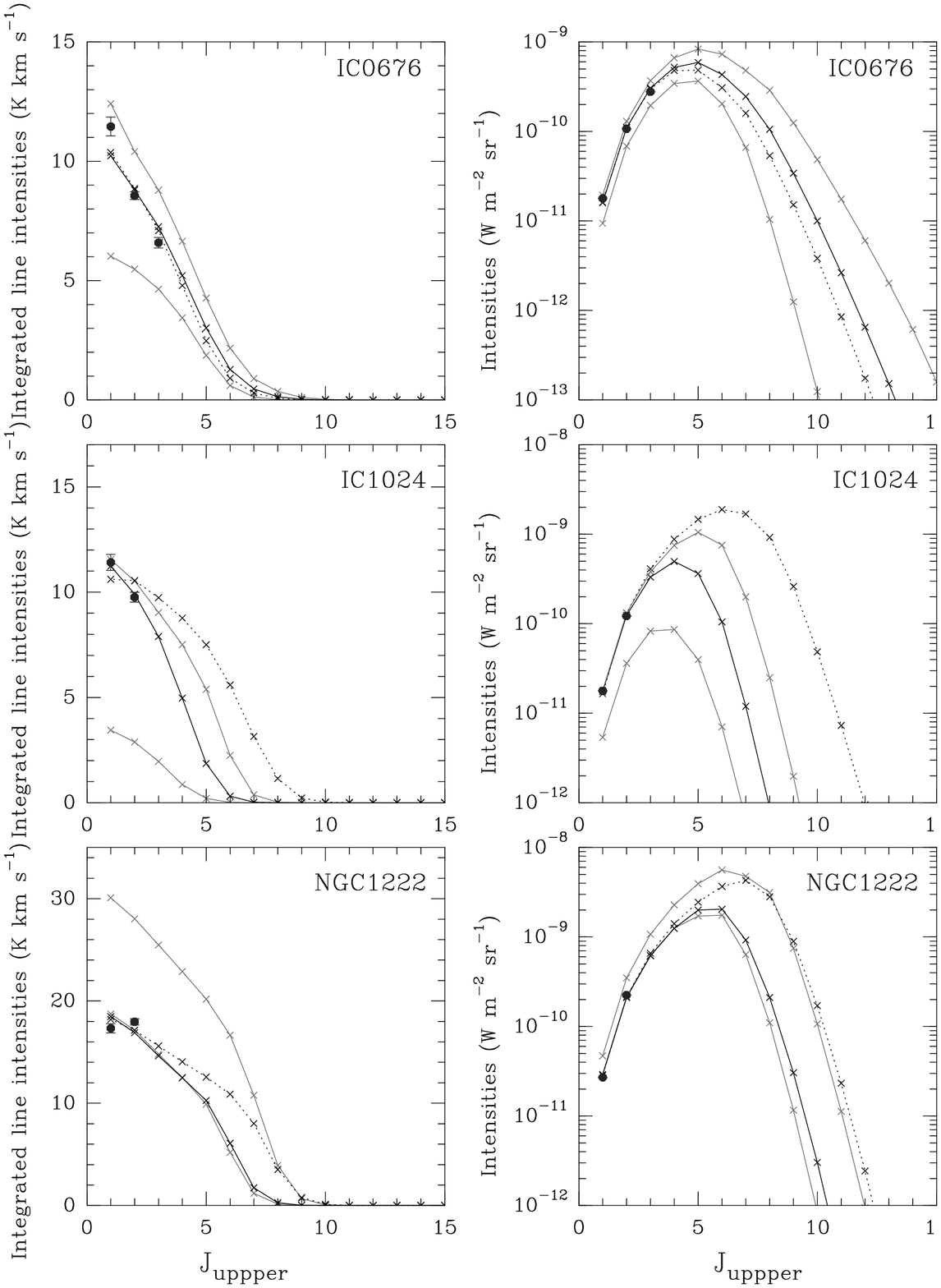}
  \caption{Predicted $^{12}$CO SLEDs. For each galaxy (indicated in
    the top-right corner of each plot), the SLED is expressed in
    K~km~s$^{-1}$ (left) and in W~m$^{-2}$~sr$^{-1}$ (right). The
    solid and dotted black lines show the predictions from the
    best-fit and most likely model, respectively. The solid grey lines
    define the range of possible SLEDs associated with the $1\sigma$
    confidence level on the best-fit model parameters (see
    Section~\ref{subsec:SED}). Crosses indicate the model
    predictions. The $^{12}$CO observations are represented by solid
    black circles with error bars.}
  \label{fig:9}
\end{figure*}

\begin{figure*}
  \addtocounter{figure}{-1}
  \centering
  \includegraphics[scale=0.8]{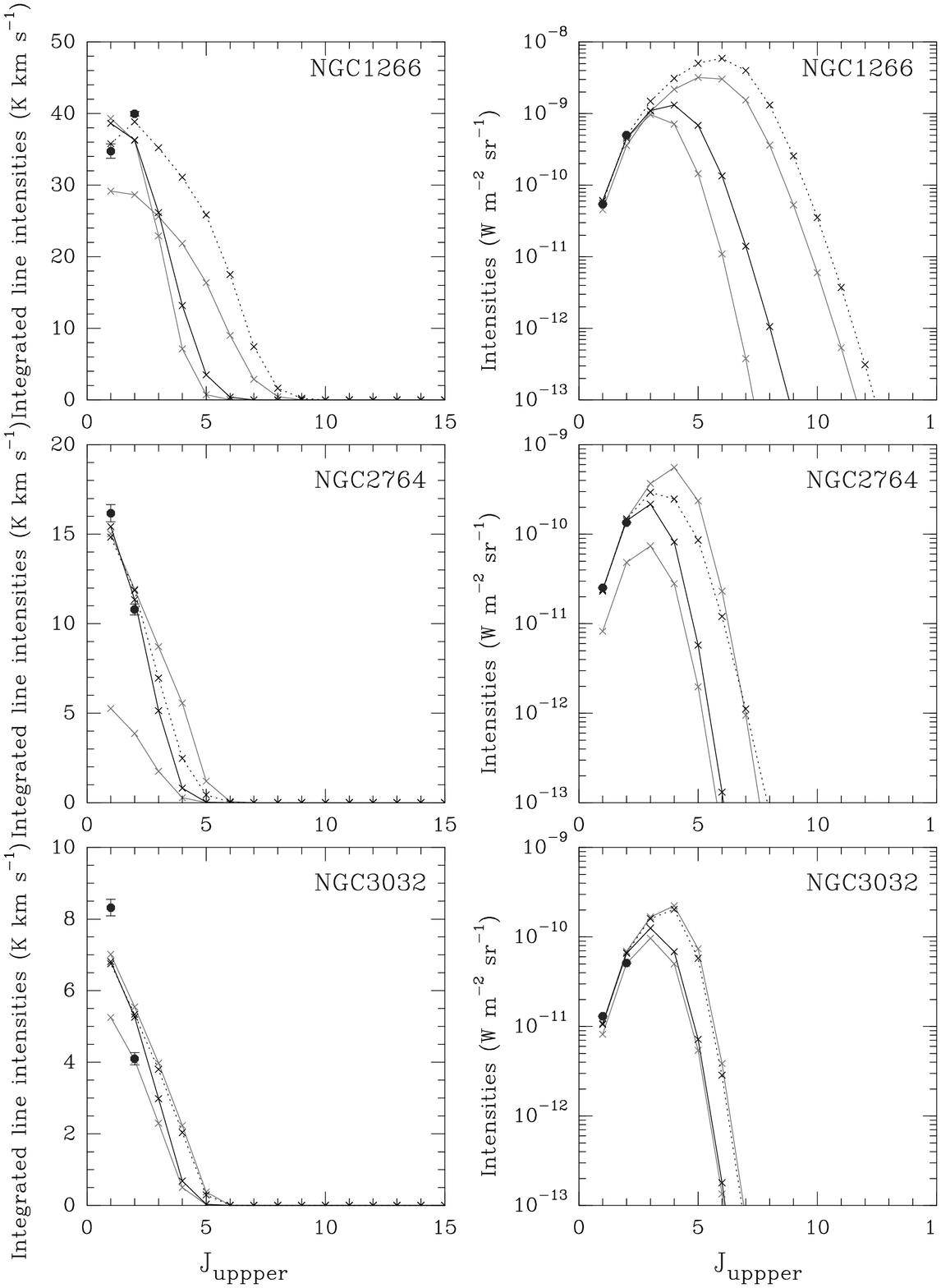}
  \caption{(Continued).}
\end{figure*}

\begin{figure*}
  \addtocounter{figure}{-1}
  \centering
  \includegraphics[scale=0.8]{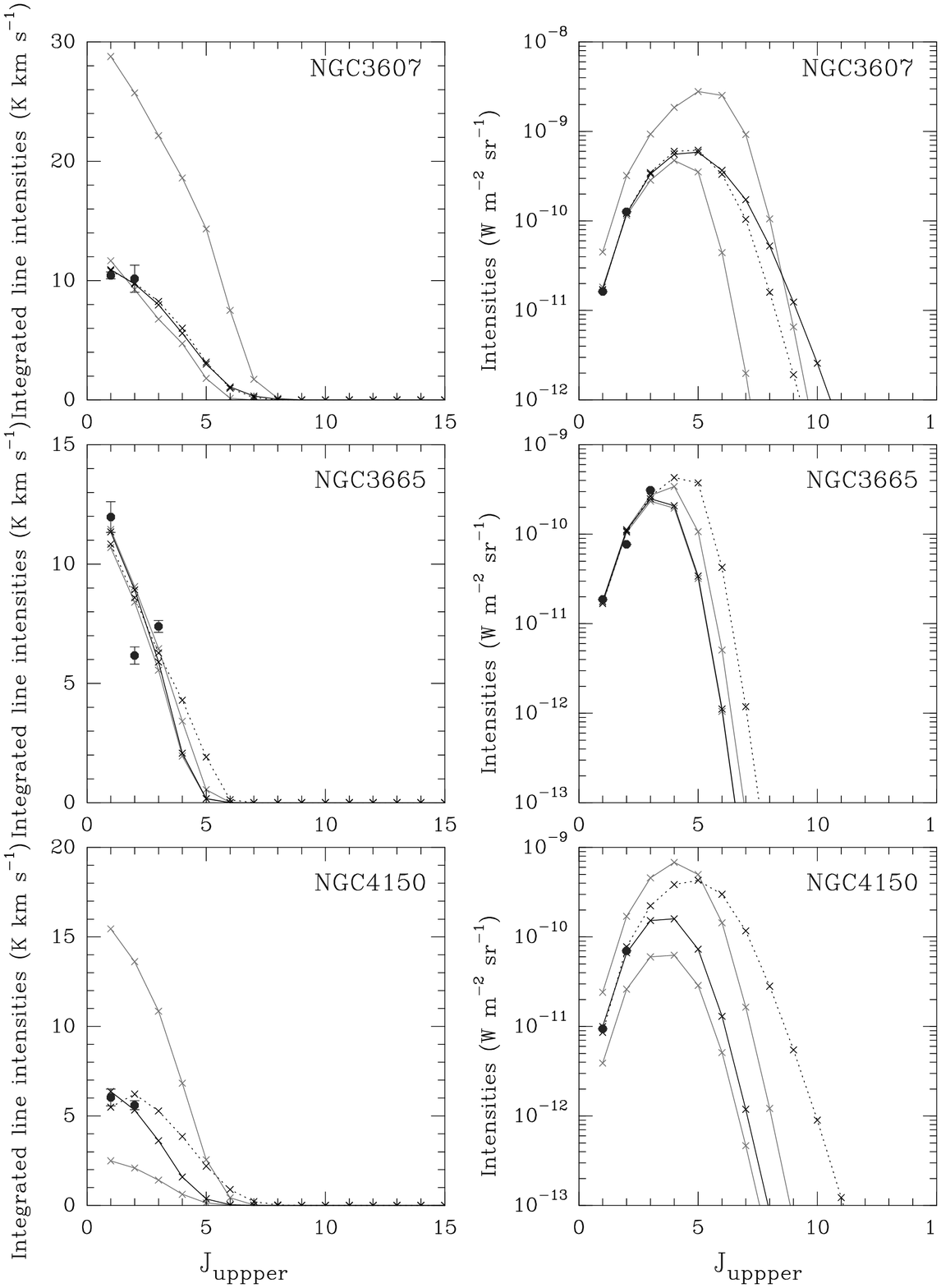}
  \caption{(Continued).}
\end{figure*}

\begin{figure*}
  \addtocounter{figure}{-1}
  \centering
  \includegraphics[scale=0.8]{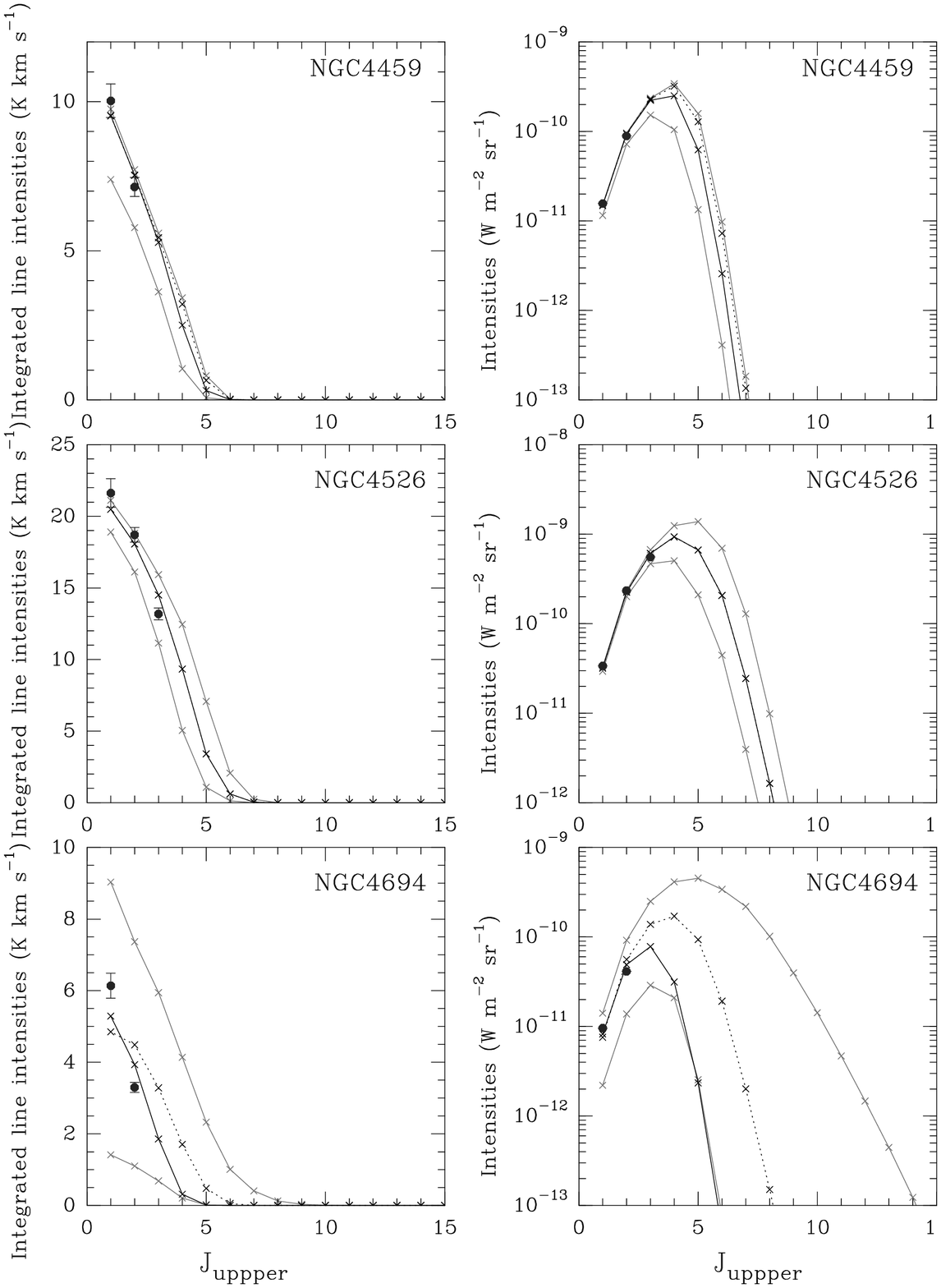}
  \caption{(Continued).}
\end{figure*}

\begin{figure*}
  \addtocounter{figure}{-1}
  \centering
  \includegraphics[scale=0.8]{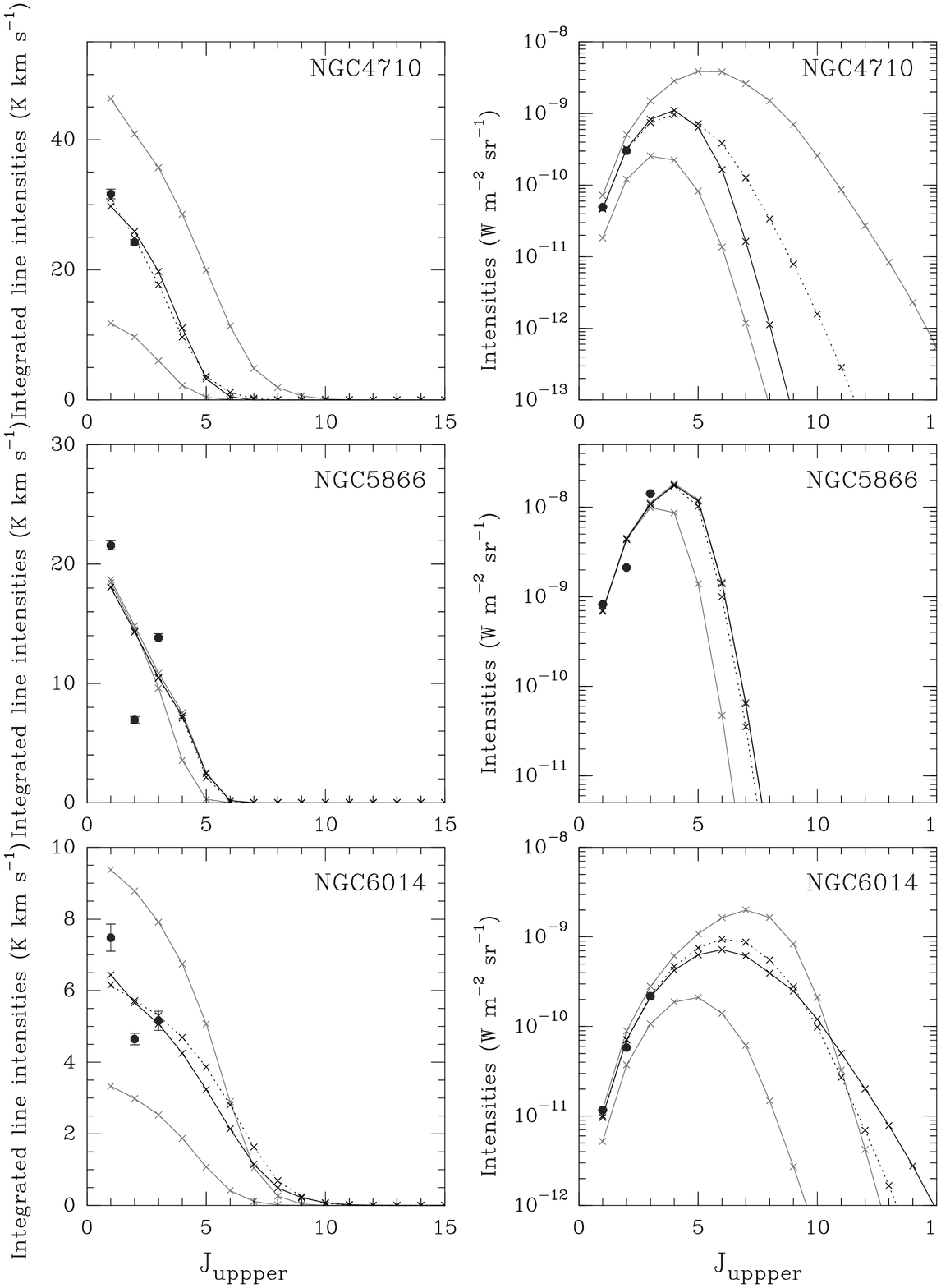}
  \caption{(Continued).}
\end{figure*}

\begin{figure*}
  \addtocounter{figure}{-1}
  \centering
  \includegraphics[scale=0.8]{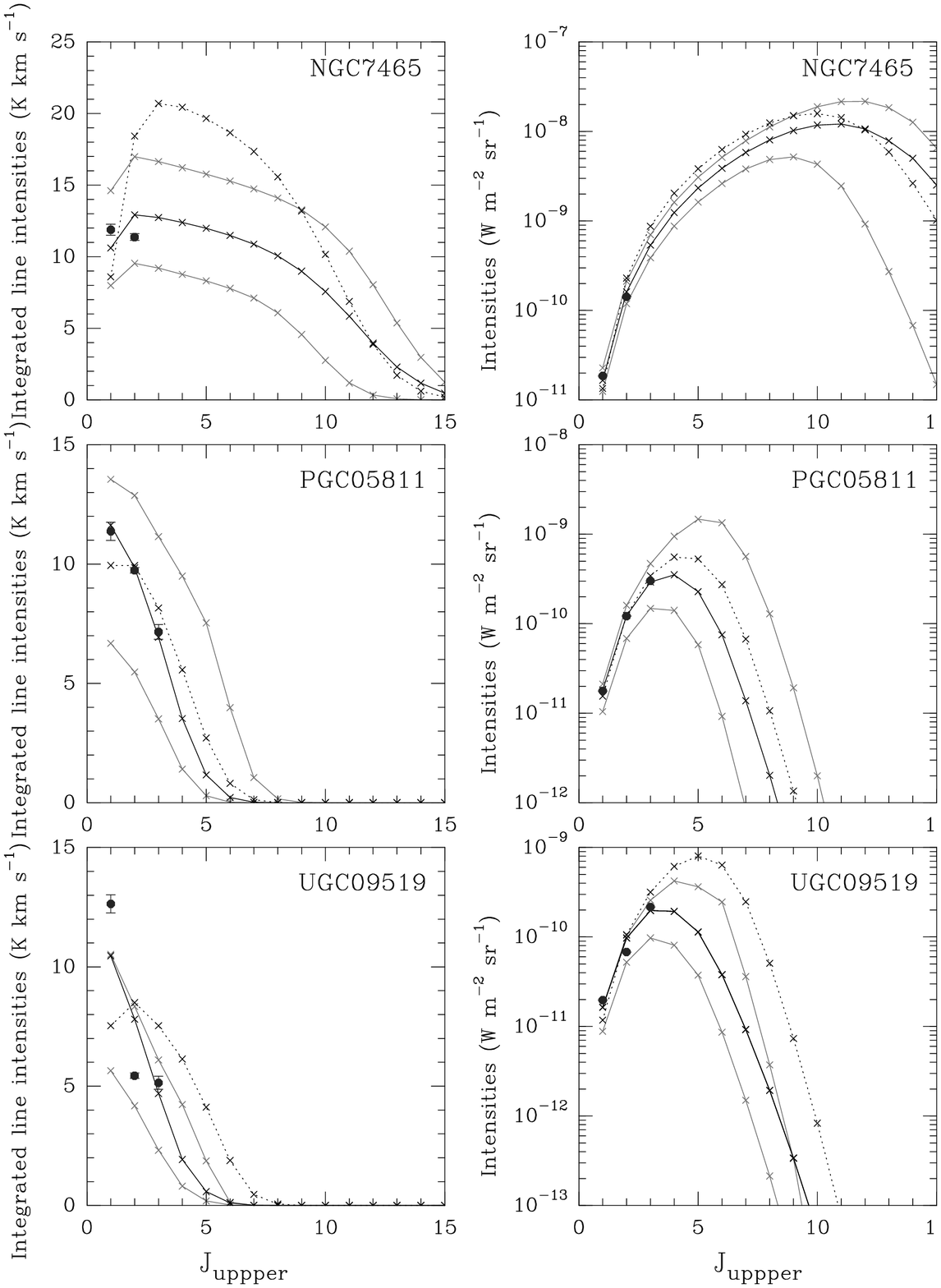}
  \caption{(Continued).}
\end{figure*}

\begin{figure*}
  \centering
  \includegraphics[scale=0.7, clip, trim=0.0cm 0.0cm 0cm 0cm]{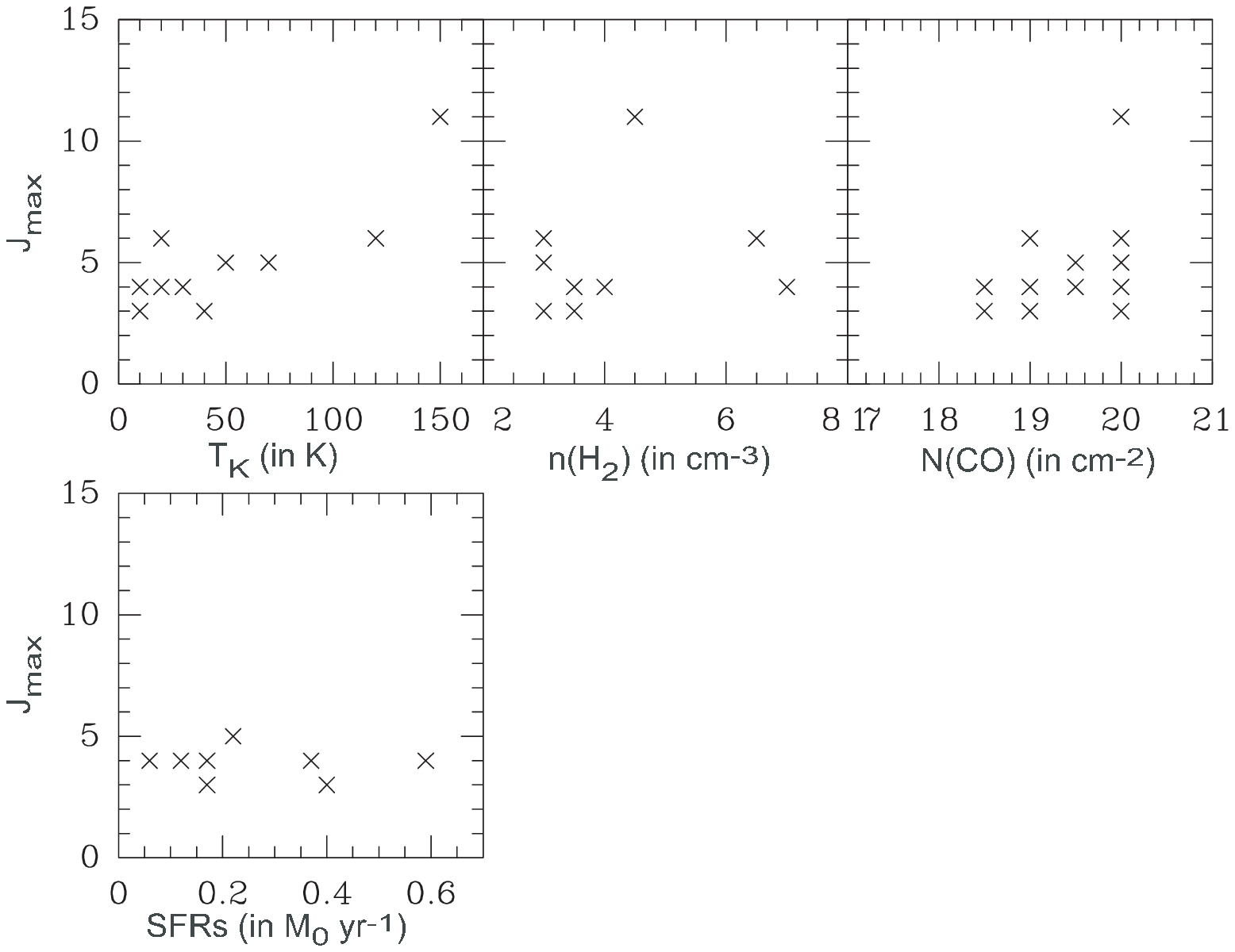}
  \caption{Top: Predicted SLED turnover position (J$_{\rm max}$) as a
    function of each of the best-fit model parameters ($T_{\rm K}$,
    $n$(H$_{2}$) and $N$(CO)). Bottom: Predicted SLED turnover
    position (J$_{\rm max}$) as a function of the $8$~$\mu$m SFRs from
    \citet{Shap10} and Falcon-Barroso et al.\ (in prep.), for the $8$
    sample galaxies where this is available (see
    Section~\ref{sec:resu}).}
  \label{fig:2}
\end{figure*}

\begin{figure*}
  \centering
  \includegraphics[scale=0.7, clip, trim=0.0cm 0.0cm 0cm 0cm]{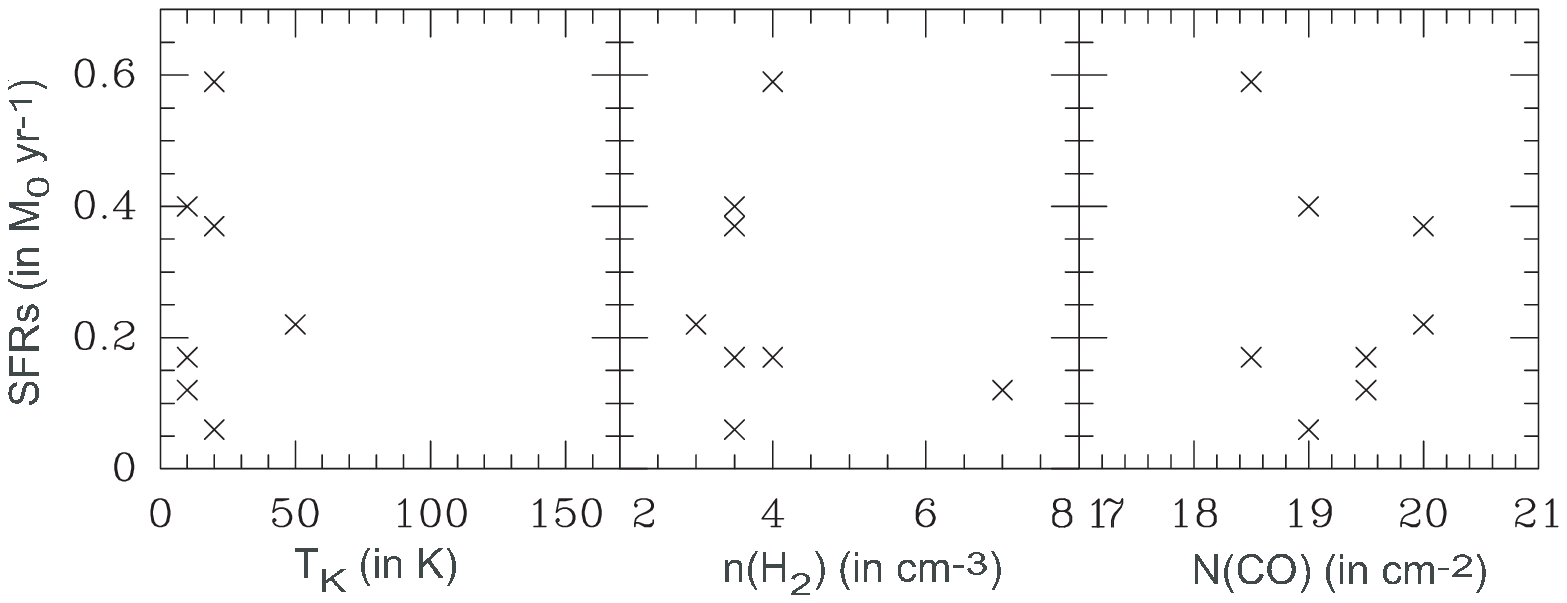}
  \caption{SFR estimates (from \citealt{Shap10} and Falcon-Barroso et
    al.\, in prep.) as a function of each of the the best-fit model
    parameters ($T_{\rm K}$, $n$(H$_{2}$) and $N$(CO)), for the $8$
    sample galaxies where this is available (see
    Section~\ref{sec:resu}).}
  \label{fig:3}
\end{figure*}

\begin{table}
  \caption{Turnover positions of the best-fit model SLEDs.}
  \label{tab:5}
  \begin{tabular}{lrr}
    \hline
    Galaxy    &  J$_{\rm max}$ & SFR \\
              &               & ($M_{\odot}$~yr$^{-1}$) \\
    \hline
    IC0676    &  $5^{+0}_{-0}$ & \\
    IC1024    &  $4^{+1}_{-0}$ & \\
    NGC1222   &  $6^{+1}_{-0}$ & \\
    NGC1266   &  $4^{+2}_{-1}$ & $0.59$ \\
    NGC2764   &  $3^{+1}_{-0}$ & \\
    NGC3032   &  $3^{+1}_{-0}$ & $0.40$ \\
    NGC3607   &  $5^{+0}_{-1}$ & $0.22$ \\
    NGC3665   &  $3^{+1}_{-0}$ & \\
    NGC4150   &  $4^{+0}_{-1}$ & $0.06$ \\
    NGC4459   &  $4^{+0}_{-1}$ & $0.17$ \\
    NGC4526   &  $4^{+1}_{-0}$ & $0.37$ \\
    NGC4694   &  $3^{+2}_{-0}$ & $0.17$ \\
    NGC4710   &  $4^{+2}_{-1}$ & \\
    NGC5866   &  $4^{+0}_{-1}$ & $0.12$ \\
    NGC6014   &  $6^{+1}_{-1}$ & \\
    NGC7465   & $11^{+1}_{-2}$ & \\
    PGC058114 &  $4^{+1}_{-1}$ & \\
    UGC09519  &  $3^{+1}_{-0}$ & \\
    \hline
  \end{tabular}

  {\bf Notes:} J$_{\rm max}$ uncertainties are defined as the
  maximum J$_{\rm max}$ ranges corresponding to a $1\sigma$ uncertainty on the
  best-fit model parameters. SFR values are from \citet{Shap10} and Falcon-Barroso et
  al.\ (in prep.).
\end{table}

\section{DISCUSSION}\label{sec:disc}

\subsection{Correlations with global parameters}\label{subsec:corrs}

{\bf In Paper~XI, ratios of molecular lines discussed in this paper
  were shown to correlate with some global physical parameters of the
  sample galaxies (e.g.\ absolute $K_{\rm s}$-band magnitude, H{\small
    I}-to-H$_{2}$ mass ratio, dust morphology, etc). It is thus
  natural to ask if the molecular gas physical conditions derived here
  show similar correlations.} We therefore investigated if the rough
groups of sources identified in Sections~\ref{subsec:chi} and
\ref{subsec:likely} (see Tables~\ref{tab:2} and \ref{tab:3}), based on
common molecular gas physical conditions, are related to other
characteristics of the galaxies derived from various other ATLAS$^{\rm 3D}$
datasets. We investigated potential correlations of our derived gas
parameters with most available quantities: distance, absolute $K_{\rm
  s}$-band magnitude, kinematic misalignment angle of the molecular
gas with respect to the stars, inclination, H{\small I} mass, H$_{2}$
mass, $60$ to $100$~$\mu$m flux ratio ($f_{60}/f_{100}$), H$\beta$
emission line equivalent width, H$\beta$, Mg$\,b$, Fe5015 and
Fe5270$_{\rm S}$ absorption line strength indices, stellar age,
metallicity ([Fe/H]) and $\alpha$-element (over-)abundance
([$\alpha$/Fe]), and effective radius (R$_{\rm e}$). Most of these
parameters were obtained from the ATLAS$^{\rm 3D}$ survey papers
(\citealt{McDe06a, McDe06b, McDe07}; Paper~I; Paper~II; Paper~III; Paper~IV; \citealt{Davi11a}, hereafter Paper~V;
Paper~VII; Paper~X; Paper~XI) or from Sarzi et al.\ (priv.\
comm.). No correlation has been identified with any of the three model
parameters ($T_{\rm K}$, $n$(H$_{2}$) and $N$(CO)) reproducing best
the observations.

A possible explanation for the absence of any correlation is that,
although the low-J CO transitions are good tracers of the total
molecular gas reservoir, only higher-J CO transitions closer to
forming stars may correlate well with, e.g., stellar properties. Or
perhaps the objects studied here simply do not span a large enough
range in stellar mass. Only future observations of higher-J CO
transitions (e.g.\ $^{12}$CO(6-5)) will confirm such a hypothesis. If
a substantial fraction of the molecular gas in ETGs has an external
origin, as argued in Paper~X, then correlations with the dominant old
stellar population may in fact not be expected. Hence, the absence of
correlations may simply support the case for an external gas
origin. {\bf However, given the existence of correlations with the
  observed line ratios (Paper~XI), the absence of correlations with
  the derived physical conditions is puzzling. It may therefore also
  suggest that the models considered here are too primitive and do not
  do justice to the likely complexity of the ISM in the sample
  galaxies. Multi-component models mixing photons dominated, X-rays
  dominated and cosmic-rays dominated regions will be investigating soon.}

\subsection{Literature comparison}\label{subsec:comp}

To further understand the star-formation activity of our molecular
gas-rich ETGs, we compared their SLEDs and best-fit model parameters
to those obtained for well-known nearby galaxies ($<10$~Mpc),
including starbursts (Maffei2, a young starburst; NGC253, an
intermediate-stage starburst; and M82, a post-starburst), a cosmic
ray-dominated galaxy (NGC6946), AGN-dominated galaxies (Markarian231
and Arp220) and normal spiral galaxies like the Milky Way. This small
sample of nearby galaxies is not exhaustive, but it rather aims to be
representative of the various types of star-formation activity present
locally. We can thus attempt to assess whether the processes at work
in these well-known `archetypes' of each star-forming class are the
same (or give rise to the same gas excitation conditions) as those in
ETGs. We know that the formation and evolution of ETGs and late-type galaxies are
very different, but are these differences also reflected in the
star-forming gas (also confined to a disc in the majority of our
sample galaxies; e.g.\ Paper~XIV; Paper~XVIII)? Only a rough
comparison is possible here because of the quality and incompleteness
of the datasets available in the literature, but it is a useful first
step.

To reproduce the low-J CO line emission of the young starburst galaxy
Maffei2, a kinetic temperature $T_{\rm K}>50$~K, an H$_{2}$ volume
density $n$(H$_{2}$)$\sim10^{3.5}$~cm$^{-3}$ and a CO column density
$N$(CO)$\approx5-7\times10^{16}$~cm$^{-2}$ are required \citep{Weli88,
  Hurt93, Dumk01}. A prediction of the CO SLED turnover position is
not available in the literature. \citet{Baye04} showed that the CO
SLED of NGC253, at an intermediate starburst evolution stage, peaks at
$^{12}$CO(6-5), whereas their best-fit model parameters are $T_{\rm
  K}=70$-$150$~K, $n$(H$_{2}$)$\gtrsim1\times10^{4}$~cm$^{-3}$ and
$N$(CO)$=1.5\pm0.5\times10^{19}$~cm$^{-2}$. Recent Herschel Observatory
results on M82, a post-starburst galaxy, show that the CO SLED peaks
at $^{12}$CO(7-6) and that $T_{\rm K}=350$-$825$~K,
$n$(H$_{2}$)$=10^{3-4.1}$~cm$^{-3}$ and
$N$(CO)$=10^{18.5-19.8}$~cm$^{-2}$ are required to reproduce the
observed CO data \citep{Panu10}. We note however that observations up
to the $^{12}$CO(13-12) transition have been obtained, and that several gas
components (whose parameters range as above) are needed to reproduce
the full SLED. The authors do not provide the gas parameters for each
component, so only the lower values of the ranges mentioned here are
appropriate for comparison with our study. \citet{Baye06} showed that
the CO gas component in the cosmic ray-dominated galaxy NGC6946
requires $T_{\rm K}=130$~K, $n$(H$_{2}$)$=1.5\times10^{3}$~cm$^{-3}$
and $N$(CO)$=4.4\times10^{14}$~cm$^{-2}$ to be excited, and that its
CO SLED peaks at $^{12}$CO(5-4). The AGN-dominated galaxy Markarian231
shows an unusual CO SLED \citep{VanderWerf10}, becoming flat after the
$^{12}$CO(6-5) transition. Several highly excited X-ray zones combined with
less excited photon-dominated regions are thought to be responsible
for this behaviour. This combination yields $T_{\rm K}=70$-$170$~K and
$n$(H$_{2}$)$=10^{3-5}$~cm$^{-3}$ globally, without indicating a
specific CO column density. Similarly to the case of M82, only the
lower portions of these ranges are relevant here, i.e.\ those
susceptible to reproduce the low-J CO transitions. Another galaxy
representing well AGN-dominated star formation activity, although
perhaps less exotic than Markarian231, is Arp220. \citet{Grev09},
without providing the CO SLED, show that the kinematic temperature
ranges from $40$ to $60$~K and the molecular gas volume density
$n$(H$_{2}$)$\approx3\times10^{2}$~cm$^{-3}$, again without indicating
a CO column density. Finally, for the center of the Milky Way, a
normal spiral galaxy, \citet{Baye04} and references therein showed
that the CO SLED peaks at $^{12}$CO(4-3), while \citet{Sand92} report
$T_{\rm K}=15$-$20$~K and $n$(H$_{2}$)$=10^{3-4}$~cm$^{-3}$, without
indicating a CO column density.

Here, despite the lack of high-J observations to confirm the shape of
our best-fit SLEDs, but assuming that the predicted SLEDs shown in
Figure~\ref{fig:9} (black solid lines) are reliable (see
Section~\ref{subsec:SED}), we see that the predicted CO SLEDs of most
of our ETGs ($13/18$) peak around the (3-2) or (4-3)
transition. Taking into account the best-fit model parameters
obtained, it is thus likely that most of our molecular gas-rich ETGs
have similar gas excitation conditions, and have a star-formation
activity similar to that present in the center of the Milky
Way. IC0676 and NGC3607 show CO SLED turnovers located around the
$^{12}$CO(5-4) transition that, when also taking into account the
best-fit model parameters ($T_{\rm K}$, $n$(H$_{2}$) and $N$(CO)) for
those sources, suggest a closer proximity to Maffei2's
characteristics. NGC1222 and NGC6014 have molecular gas and CO SLED
properties closer to those of NGC253, while NGC7465 seems in agreement
with Markarian231. Again, however, although these reults are
suggestive and interesting, we must be careful to not over-interpret
them as they are extrapolations from relatively low-J CO lines
only. With additional higher-J CO transition data, more precise
comparisons and confirmation of these results will be possible.

\section{Conclusions}\label{sec:con}

Exploiting recently published molecular line ratios, we have modelled
for the first time the CO and HCN-HCO$^{+}$ gas components in a
sizeable sample of gas-rich early-type galaxies ($18$), this using a
non-LTE theoretical method (i.e.\ an LVG code). We have separated
these two gas components as the molecules studied are known to probe
different gas phases, hence a priori should trace different physical
conditions. While the HCN and HCO$^{+}$ lines alone leave the
modelling of their corresponding gas phase degenerate, we are able to
properly constrain the gas kinetic temperature, H$_{2}$ volume density
and CO column density of the CO gas component using the $^{12}$CO(1-0,
2-1, 3-2) and $^{12}$CO(1-0, 2-1) transitions.

Most galaxies in our sample have a relatively cool ($10$-$20$~K) gas,
a common H$_{2}$ volume density ($n$(H$_{2}$)$\sim10^{3-4}$~cm$^{-3}$)
and an average CO column density ($N$(CO)$>10^{18}$~cm$^{-2}$). For
the first time, we also provide the predicted CO spectral line energy
distributions of these $18$ ETGs, including the predicted SLED peak
position. Taking into account the gas excitation conditions, derived
physical properties, and the model assumptions and limitations, most
of the molecular gas in our gas-rich ETGs appears to host
star-formation activity at a level similar to that present in the
center of the Milky Way. Since their CO SLEDs show higher excitation
conditions, a few galaxies (IC0676, NGC1222, NGC3607, NGC6014 and NGC7465) may
have slightly more active star formation or additional sources of
excitation (AGN, cosmic ray-dominated regions, etc). 
Strong $24$, $60$ and $160$~$\mu$m {\it Spitzer} fluxes in NGC3607
(compared to average values for ETGs) seem to confirm this hypothesis
\citep{Temi09}.

The data used to constrain our models are averaged over several
kiloparsecs and often over the entire galaxy. Local (sub-kpc) gas
properties could thus be significantly different from the estimates
presented here, and the values listed in Table~\ref{tab:2} and
\ref{tab:3} should be used with caution. Nevertheless, these values
are the best estimates of the molecular gas properties of ETGs
available, and will remain so until spatially-resolved interferometric
data of multiple species and transitions become available.

This is also the first time that the physical properties are derived
for two separate cold gas components in ETGs. This is a huge step
forward in our understanding of current (residual) star formation in
these objects. Despite the large uncertainties in the results derived
for the HCN and HCO$^{+}$ gas component, our study nevertheless
clearly shows that the physical properties of this component differ
from those of the CO gas component, even in ETGs. This confirms the
need for a separate treatment of these two gas components, with
possibly different evolutionary paths.

It is also essential to identify additional transitions of CO, HCN and
HCO$^{+}$ likely to improve these results and better constrain the
physical properties of the dense gas in ETGs in the future. We have
identified here the mid-J CO transitions (e.g.\ $^{12}$CO(6-5)) as
particularly useful, with much constraining power.

\section*{Acknowledgments}

The authors thank Dr.\ S.\ Kraviraj for useful discussions on the
likelihood method and the referee for his/her useful comments which
greatly improved the paper. The authors acknowledge financial support
from ESO. EB acknowledges financial support from STFC rolling
grant `Astrophysics at Oxford 2010-2015' (ST/H002456/1) and John Fell
OUP Research Fund "Molecules in galaxies: securing Oxford's position
in the ALMA era (092/267)". TAD acknowledges funding from the European Community's Seventh Framework
Programme (/FP7/2007-2013/) under grant agreement No 229517. LMY acknowledges financial support from NSF
grant 1109803. MBois and TN acknowledge support from the
DFG Cluster of Excellence 'Origin and Structure of the Universe'. MBois has received,
during this research, funding from the European Research Council under
the Advanced Grant Program Num 267399-Momentum. MC
acknowledges support from a Royal Society University Research
Fellowship. RLD acknowledges travel and computer grants from
Christ Church, Oxford and support from the Royal Society in the form
of a Wolfson Merit Award 502011.K502/jd. RLD also acknowledges the
support of the ESO Visitor Programme which funded a 3 month stay in
2010. SK acknowledges support from the the Royal Society Joint
Projects Grant JP0869822. RMcD is supported by the Gemini Observatory, which is
operated by the Association of Universities for Research in Astronomy,
Inc., on behalf of the international Gemini partnership of Argentina,
Australia, Brazil, Canada, Chile, the United Kingdom, and the United
States of America. MS acknowledges support from a STFC Advanced
Fellowship ST/F009186/1. PS is a NWO/Veni fellow.

\newcommand{\apj}[1]{ApJ, }
\newcommand{\apss}[1]{Ap\&SS, }
\newcommand{\aj}[1]{Aj, }
\newcommand{\apjs}[1]{ApJS, }
\newcommand{\apjl}[1]{ApJ Letter, }
\newcommand{\aap}[1]{A\&A, }
\newcommand{\aaps}[1]{A\&A Suppl. Series, }
\newcommand{\araa}[1]{Annu. Rev. A\&A, }
\newcommand{\aaas}[1]{A\&AS, }
\newcommand{\bain}[1]{Bul. of the Astron. Inst. of the Netherland,}
\newcommand{\mnras}[1]{MNRAS, }
\newcommand{\nat}[1]{Nature, }
\newcommand{\araaa}[1]{ARA\&A, }
\newcommand{\planss}[1]{Planet Space Sci., }
\newcommand{\jrasc}[1]{Jr\&sci, }
\newcommand{\pasj}[1]{PASJ, }
\newcommand{\pasp}[1]{PASP, }
\newcommand{\qjras}[1]{QJRAS, }
\newcommand{\nar}[1]{NewAR,}
\bibliographystyle{mn2e_2}
\bibliography{references}

\appendix

\section{RESULTS FOR THE HCN AND HCO$^{+}$ GAS
  COMPONENT}\label{app:fig}

\begin{table}
  \caption{Best-fit model parameters of the HCN and HCO$^{+}$ gas
    component.} \label{tab:A1}
  \begin{tabular}{lrrrr}
    \hline
    Galaxy & $\chi^{2}$ & $T_{\rm K}$ & $n$(H$_{2}$) & $N$(HCN,HCO$^{+}$)\\
           &            & (K)        & (cm$^{-3}$)  & (cm$^{-2}$) \\
    \hline
    IC0676  & $2.91\times10^{-3}$ & 140 & $10^{6\phantom{.0}}$ & $10^{19.5}$\\
    NGC1266 & $6.96\times10^{-4}$ &  90 & $10^{6.5}$          & $10^{19.5}$\\
    NGC2764 & $6.50\times10^{-6}$ & 190 & $10^{4\phantom{.0}}$ & $10^{19\phantom{.0}}$\\
    NGC3607 & $0.10\phantom{x10^{-0}\,\,\,\,}$ & 80 & $10^{6\phantom{.0}}$ & $10^{18.5}$\\
    NGC4710 & $4.72\times10^{-2}$ &  30 & $10^{3\phantom{.0}}$ & $10^{20\phantom{.0}}$\\
    NGC5866 & $4.64\times10^{-2}$ &  10 & $10^{3\phantom{.0}}$ & $10^{20\phantom{.0}}$\\
    NGC7465 & $3.14\times10^{-5}$ & 170 & $10^{5\phantom{.0}}$ & $10^{16\phantom{.0}}$\\
    \hline
  \end{tabular}
\end{table} 

\begin{figure*}
  \begin{minipage}[c]{.30\linewidth}
    \includegraphics[scale=0.8, clip, trim=0cm 0.5cm 0cm 0cm]{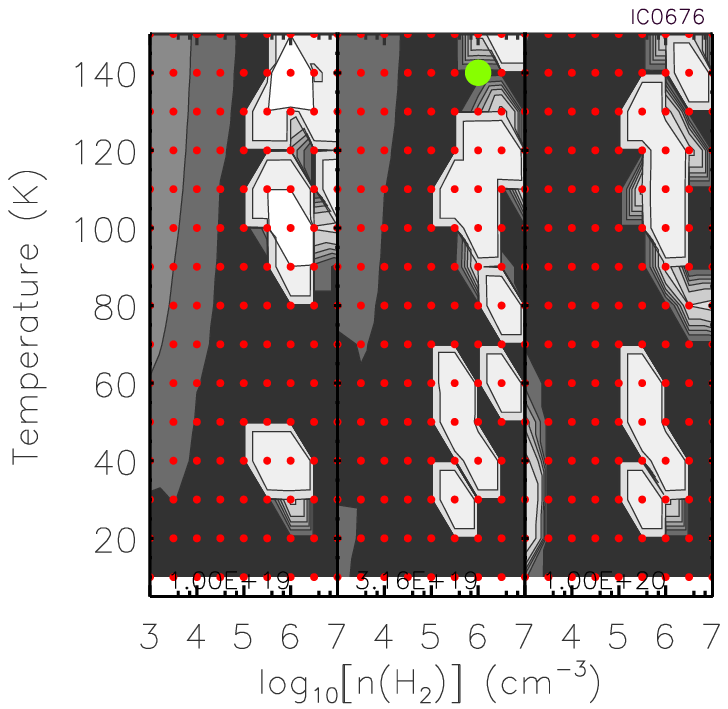}
  \end{minipage}
  \vspace*{-0.6cm} \hspace*{0.6cm}
  \begin{minipage}[c]{.30\linewidth}
    \includegraphics[scale=0.80, clip, trim=0.8cm 0.5cm 0cm 0cm]{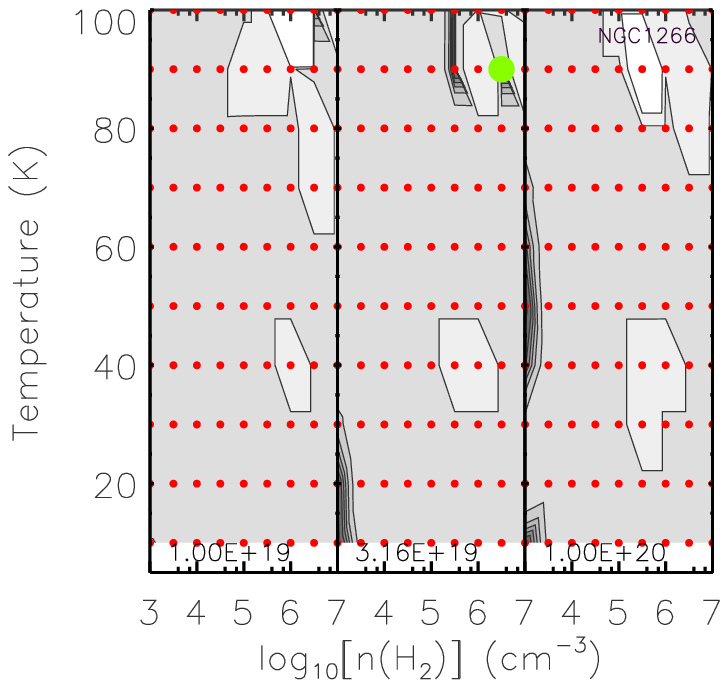}
  \end{minipage}
  \begin{minipage}[c]{.30\linewidth}
    \includegraphics[scale=0.80, clip, trim=0.75cm 0.5cm 0cm 0cm]{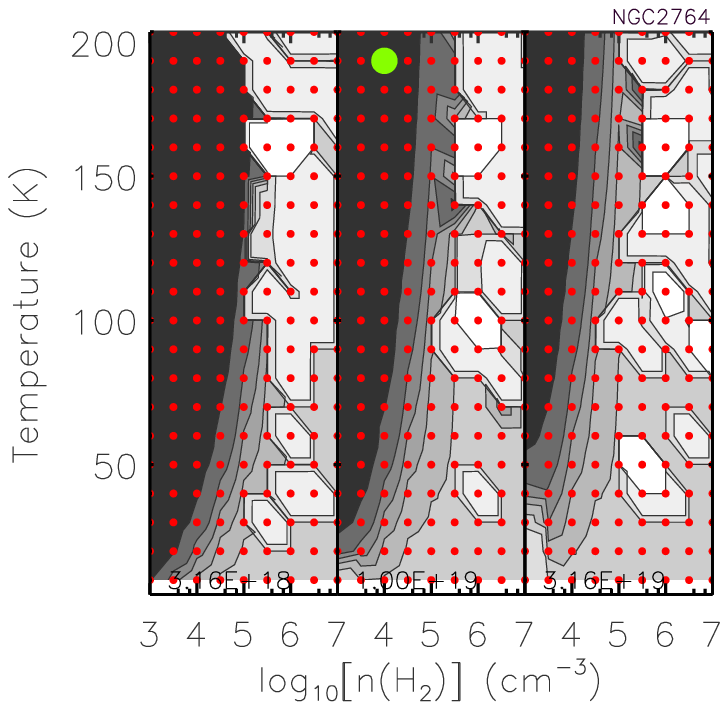}
  \end{minipage}
  \vspace*{-0.6cm}
  \begin{minipage}[c]{.30\linewidth}
    \includegraphics[scale=0.80, clip, trim=0cm 0cm 0cm 0cm]{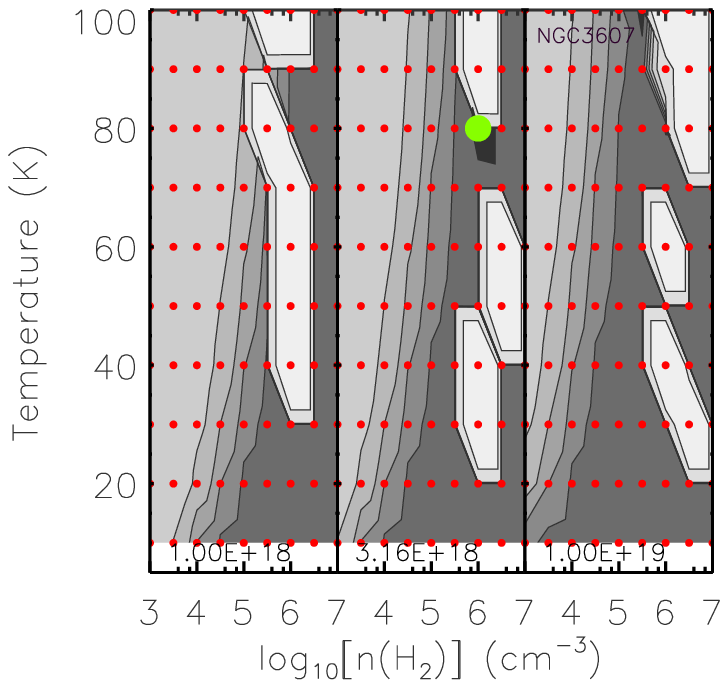}
  \end{minipage}
  \hspace*{0.9cm}
  \begin{minipage}[c]{.30\linewidth}
    \includegraphics[scale=0.80, clip, trim=0.9cm 0cm 0cm 0cm]{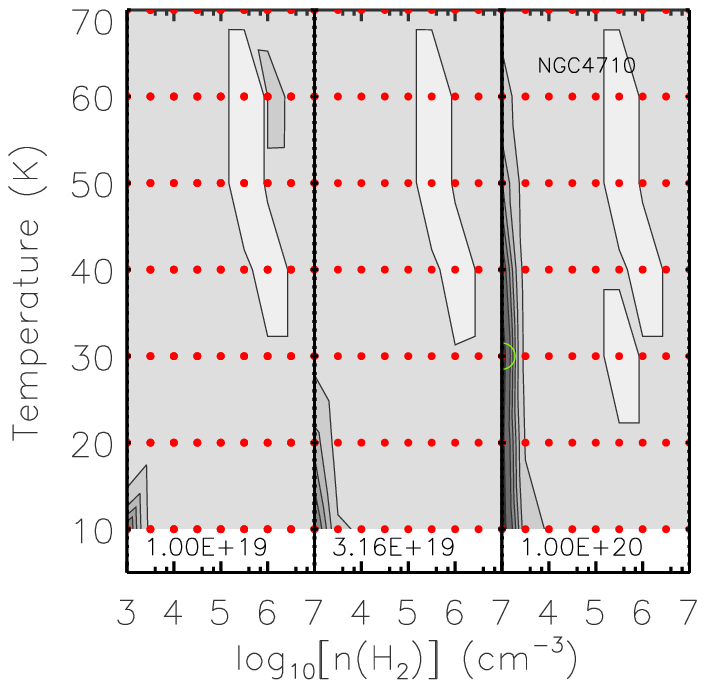}
  \end{minipage}
  \begin{minipage}[c]{.30\linewidth}
    \includegraphics[scale=0.80, clip, trim=0.9cm 0cm 0cm 0cm]{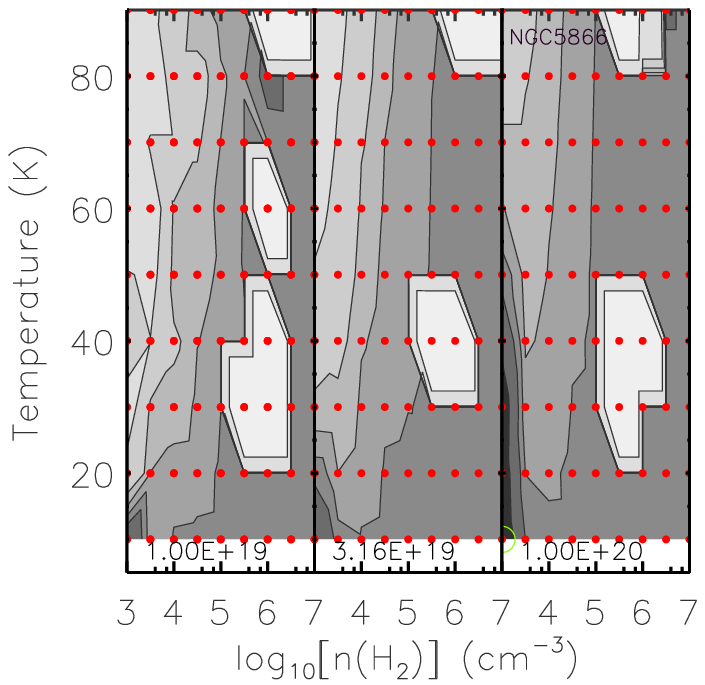}
  \end{minipage}
  \begin{minipage}[l]{.30\linewidth}
    \includegraphics[scale=0.80, clip, trim=0cm 0cm 0cm 0cm]{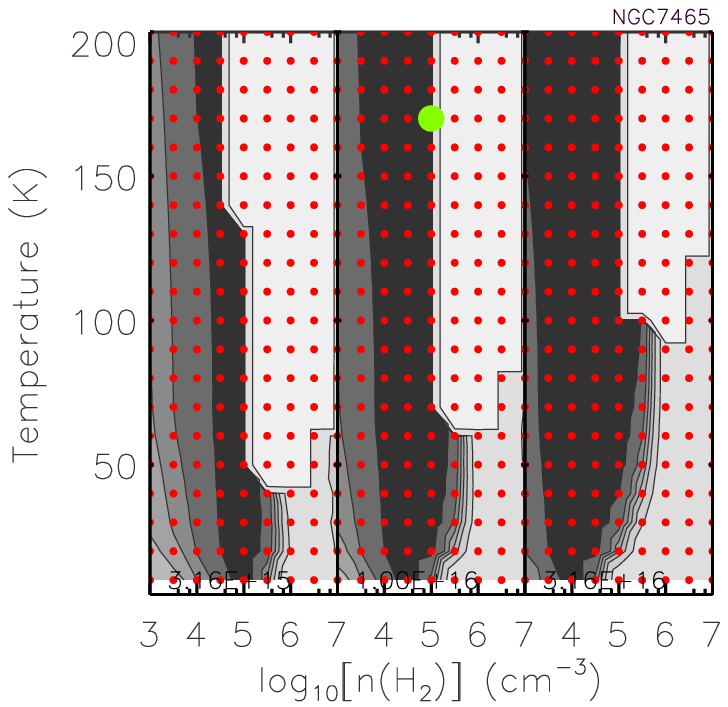}
  \end{minipage}
  \caption{Best-fit models for the HCN and HCO$^{+}$ gas
    component. Each panel shows the
    $\Delta\chi^2\equiv\chi^2-\chi^2_{\rm min}$ contours of the HCN
    and HCO$^{+}$ gas component as a function of the kinetic
    temperature $T_{\rm K}$ and H$_{2}$ volume density $n$(H$_{2}$),
    for three values of the CO column density $N$(CO) (hence three
    plots) centered on the best-fit value and indicated at the bottom
    of each plot. The model grid is shown with dark-grey dots (red
    dots in the online version), whereas the best-fit model of each
    galaxy ($\chi^{2}_{\rm min}$ or $\Delta\chi^{2}=0$) is labeled
    with a light-grey filled circle (green filled circle in the online
    version). The $\Delta\chi^2$ contours and greyscales are typically
    for $1$ to $5\sigma$ confidence levels (i.e.\ $\Delta\chi^2=1.0$,
    $2.7$, $3.8$, $5.0$ and $6.6$). The low $\sigma$, high confidence
    level contours containing 'good' models are represented by the
    darkest areas. White zones correspond to models where there is no
    solution for fitting the single line ratio
    HCN(1-0)/HCO$^{+}$(1-0). These models produce either negative
    opacities or negative integrated line intensities (see limitations
    in \citealt{VanderTak07}). The galaxy name is indicated in the
    top-right corner of each panel.}
\label{fig:app_1}
\end{figure*}

We present here the modeling results for the HCN and HCO$^{+}$ gas
component, for the galaxies where both lines are
detected. Figure~\ref{fig:app_1} shows the $\Delta\chi^{2}$ contours
of the HCN-HCO$^{+}$ models, similarly to Figure~\ref{fig:1} for the
CO gas component. Table~\ref{tab:A1} lists the best-fit model
parameters. The HCN-HCO$^{+}$ results show greater uncertainties as
the models are underconstrained (one line ratio but three model
parameters). They should thus be considered indicative only and should
be used with caution.

\bsp

\label{lastpage}

\end{document}